\documentclass[aps,onecolumn,prd,superscriptaddress,nofootinbib]{revtex4}
\usepackage{epsfig}
\usepackage{axodraw}
\usepackage{amssymb}
\usepackage{amsmath}
\usepackage{amsfonts}
\newcommand{\beq}{\begin{equation}}
\newcommand{\eeq}{\end{equation}}
\newcommand{\beqn}{\begin{eqnarray}}
\newcommand{\eeqn}{\end{eqnarray}}

\begin{document}
\sloppy
\preprint{LYCEN-2006-19 \\SPhT-T06/146}
\title{Quantum corrections to the effective neutrino mass operator in 5D MSSM}
\author{Aldo Deandrea}
\affiliation{Universit\'e de Lyon 1, Institut de Physique Nucl\'eaire,
4 rue E.~Fermi, 69622 Villeurbanne Cedex, France} 
\author{Pierre Hosteins}
\affiliation{SPhT, CEA-Saclay,91191 Gif-sur-Yvette cedex, France}
\author{Micaela Oertel}
\affiliation{LUTH, Observatoire de Paris-Meudon, 5 place
  Jules Janssen, 92195 Meudon, France}
\author{Julien Welzel}
\affiliation{Universit\'e de Lyon 1, Institut de Physique Nucl\'eaire,
4 rue E.~Fermi, 69622 Villeurbanne Cedex, France}

\begin{abstract}We discuss in detail a five-dimensional Minimal Supersymmetric
Standard Model compactified on $S^1/Z_2$ extended by the effective
Majorana neutrino mass operator. We study the evolution of neutrino masses 
and mixings. Masses and angles, in particular the atmospheric mixing angle 
$\theta_{23}$, can be significantly lowered at high energies with respect 
to their value at low energy.
\end{abstract}

\maketitle

\section{Introduction}

Within the Standard Model (SM), the masses of the quarks and charged
leptons are determined {\it via} Yukawa couplings to the Higgs boson.
The origin of their structure (masses and mixing angles between
flavours) has no explanation in the context of the SM and it is one of
the major challenges for physics beyond the SM. Neutrinos being massless
in the SM, the experimental evidence for nonzero neutrino masses gives
an important indication for physics beyond the SM.  In any case,
neutrino masses are many orders of magnitude smaller than those of
quarks and charged leptons.  Moreover, the experimental results
indicate that the lepton mixing matrix has two large mixing angles
unlike the quark mixing matrix.  All this shows that the neutrino
sector plays a special role in understanding the flavour structure of
the SM and its possible extensions. 

Phenomenological implications of quantum corrections to the neutrino
mass and mixing parameters have been investigated intensively in the
literature (see e.g. the review~\cite{Chankowski:2001mx}). The main
reason is that large effects can provide interesting hints for model
building.  From a theoretical point of view, many models are available
proposing different explanations for the particularities of the
neutrino sector.  In order to study qualitative features in a model
independent way, an attractive and simple possibility is to stick to a
low-energy effective theory formulation. This means that one organises
the effects of additional particles and symmetries present at higher
energies within a systematic low-energy expansion. We will assume here
that the heavy states arising from physics beyond the standard model
completely decouple at low energies. In that case, the
degree of suppression of an operator in the low energy effective
Lagrangian is characterised by its mass dimension (d). The only
operator appearing at dimension d = 5 is the lepton-number violating
operator~\cite{BW86}
\beq 
-\frac{\kappa_{ij}'}{4
  \, M}(L^i_a\epsilon^{ab} \phi_b) (L^j_c\epsilon^{cd}\phi_d)
+\,h.c~,
\label{EffectiveOperatorSM}
\eeq
where $L$ and $\phi$ represent the lepton and the Higgs doublet
fields, respectively. $M$ is an energy scale characteristic for
the range of validity of the low-energy effective theory description.
An operator of this type can be generated, for instance, by the usual
seesaw mechanism~\cite{seesawI}.  In that case the scale
$M$ can be identified with the mass of the heavy right-handed
neutrino. After spontaneous breakdown
of the electroweak symmetry, the Higgs acquires a vacuum expectation
value ($vev$) and the operator in Eq.~(\ref{EffectiveOperatorSM}) then
represents a Majorana mass term for the neutrinos.

In the context of the Minimal Supersymmetric Standard Model
(MSSM), it can be written in the form :
\beq
-\frac{\kappa_{ij}'}{4\, M}(L^i_a\epsilon^{ab}H^{(u)}_b)(L^j_c\epsilon^{cd}H^{(u)}_d)~,
\label{EffectiveOperatorMSSM}
\eeq
where $L$  and $H^{(u)}$
now stand for the lepton and up-type Higgs doublet chiral superfield,
respectively. 
This dimension five operator
provides a very efficient way to study neutrino masses and mixings.
Renormalisation group equations for this effective operator have been
derived in
the context of the four-dimensional SM and
MSSM in Refs.~\cite{RGEkappa,Antuschkappa}.  

Scenarios with compactified extra-dimensions offer many possibilities
for model building. For example, there are new ways to generate
electroweak symmetry breaking or supersymmetry breaking simply by
choosing appropriate boundary conditions (for a review on
extra-dimensions and their phenomenology, see~\cite{XDrev}). In
addition, for flat extra-dimensions the presence of towers of excited
Kaluza-Klein states induces the power-law enhancement of the gauge
couplings, leading to a possible low-scale
unification~\cite{DDG,TeVstrings}. This effect can be applied to
other couplings such as Yukawa couplings, too, giving an original way to
generate mass hierarchies~\cite{DDG,YukawaXD}. For the same reasons,
extra-dimensions can also provide a possible explanation of the
observed pattern of neutrino masses and mixings.

The aim of this paper is to study these effects explicitly in the case
of one extra-dimension within a supersymmetric model supplemented by the
effective neutrino mass operator, Eq.~(\ref{EffectiveOperatorMSSM}). In the
following we shall consider the effects of renormalisation at one loop
in order to test the behaviour of the extra-dimensional model.  We
shall focus on a five dimensional $\mathcal{N}=1$ supersymmetric
model compactified on the orbifold $S^1/Z_2$ as a simple test ground
for the effects of the extra dimension. 

The experimental results will be 
used as a starting point
for the evolution of the masses and couplings in order to test the
evolution at higher energies and the effects induced by the presence
of the extra dimension. Due to the power-law running of the (gauge) couplings,
there are of course restrictions on the range of validity of the present 
model which consequently put limits on the present investigation. 

Instead of starting from the observed masses and mixing parameters at low 
energies, we could take the renormalisation group equations provided here, to
constrain parameters of some specific model at high energies by studying the 
evolution to low energies and comparing the predictions with data. Since we 
are mainly interested in the qualitative effect
of the extra dimension we refrain from adding assumptions on
the high energy behaviour, although this would certainly be interesting 
from the theoretical point of view. 

The paper is organised as follows. 
In section 2, after a short introduction on generic $\mathcal{N}=1$ 
supersymmetric 5D models, we present the features of a five-dimensional MSSM 
compactified on the orbifold $S^1/Z_2$ and discuss its low-energy spectrum. The details of the
Lagrangian and its Feynman rules are given in appendix 1. 
The third section is devoted to a discussion of the beta functions for the 
Yukawa couplings and the effective neutrino mass operator. 
Numerical results for the evolution of neutrino masses and
mixings are given in section 4. In the last section we summarise and 
discuss the physical implications.

\section{5D MSSM}
  \label{model}
\subsection{Five-dimensional $\mathcal{N}=1$ supersymmetry}
The beta functions can be most elegantly derived in the superfield 
formalism. We will therefore begin with briefly discussing  
$\mathcal{N}=1$ supersymmetry in a
five-dimensional Minkowski space and its description in terms of 4D 
superfields. More details can be found in 
Refs.~\cite{Flacke, Hebecker,XDSUSY}. Space-time coordinates will be denoted 
by $(x^{\mu}, y)$.

\subsubsection{Gauge sector}
The gauge sector will be described by a 5D $\mathcal{N}=1$ vector
supermultiplet which consists (on-shell) of a 5D vector field $A^M$, a real 
scalar $S$ and two gauginos, $\lambda$ and $\lambda'$.
\begin{eqnarray*}
S_g &=& \int
\mathrm{d}^5x\frac{1}{2kg^2}\mathrm{Tr}\left[-\frac{1}{2}F^{MN}F_{MN}-D^MSD_MS-i\overline{\lambda}\Gamma^MD_M\lambda
\right.\nonumber\\
&+&-i\overline{\lambda}'\Gamma^MD_M\lambda'+\left.(\overline{\lambda}+\overline{\lambda}')[S,\lambda+\lambda']\right]
\label{5DGaugeaction}
\end{eqnarray*}
with $D_M=\partial_M+iA_M$, $\Gamma^5=i\gamma^5$ and $F^{MN}=\frac{-i}{g}[D^M,D^N]$. $k$ normalises the trace over the generators of the gauge groups. 
Equivalently, one can rearrange these fields in terms of a $\mathcal{N}=2$, 
$D=4$ vector supermultiplet, $\Omega=(V\,,\chi)$, where:
\begin{itemize}
\item $V$ is a $\mathcal{N}=1$ vector supermultiplet containing $A^{\mu}$ and
$\lambda$,
\item $\chi$ is a chiral $\mathcal{N}=1$ supermultiplet containing $\lambda'$
and $S'=S+iA^5$.
\end{itemize}
This follows from the decomposition of the 5D supercharge (which is a Dirac
spinor) into two Majorana-type supercharges which constitute a $\mathcal{N}=2$
superalgebra in 4D. Both $V$ and
$\chi$ (and their component fields) are in the adjoint representation of the
gauge group $\mathcal{G}$. Using these supermultiplets, one can rewrite the
original 5D $\mathcal{N}=1$ supersymmetric action~(\ref{5DGaugeaction}) only 
in terms of $\mathcal{N}=1$ $D=4$ superfields and the covariant derivative in 
the $y$ direction~\cite{Hebecker}:
\beq
S_g = \int\mathrm{d}^5x\mathrm{d}^2\theta\mathrm{d}^2\overline{\theta}\frac{1}{4kg^2}\mathrm{Tr}
\left[\frac{1}{4}(W^{\alpha}W_\alpha\delta(\overline{\theta}^2)+\,h.c)+(e^{-2gV}\nabla_ye^{2gV})^2\right]
\eeq
with $W^{\alpha}=-\frac{1}{4}\overline{D}^2e^{-2gV}D_{\alpha} e^{2gV}$. 
$D_{\alpha}$ is the covariant derivative in the 4D $\mathcal{N}=1$ superspace 
(see the textbooks~\cite{Westbook, WessBagger}) and $\nabla_y=\partial_y+\chi$.
This action can be expanded and quantised to find the Feynman rules to a given
order in the gauge coupling $g$ .

\subsubsection{Matter sector and its coupling to the gauge sector}
The $\mathcal{N}=1$ supersymmetric and 5D Lorentz invariant action describing 
a free chiral supermultiplet is:
\beq
S = \int\mathrm{d}^5x\left((\partial_M\phi_i)^{\dag}(\partial^M\phi^i)-\overline{\psi}(i\Gamma^M\partial_M+m)\psi\right)
\label{Action5Dmatter}
\eeq
The two complex scalars $\phi^{1,2}$ form a doublet under a `$SU(2)_R$' 
symmetry and 
$\psi$ is a $SU(2)_R$-singlet Dirac spinor. Together, they can form two 4D 
$\mathcal{N}=1$ chiral supermultiplets, $\Phi$ and $\Phi^c$. 
Adding the couplings to the gauge sector we obtain for the action in terms of
the 4D superfields:
\beq
S = \int\mathrm{d}^5x\mathrm{d}^2\theta\mathrm{d}^2\overline{\theta}
\left(\overline{\Phi}e^{2gV}\Phi+\Phi^ce^{-2gV}\overline{\Phi}^c+(\Phi^c(\nabla_5+m)\Phi\delta(\overline{\theta}^2)+h.c)\right)
\eeq
This way of writing the action has the disadvantage that the 5D Lorentz 
invariance and the 
underlying supersymmetry relating $\Phi$ to $\Phi^c$ and $V$ to $\chi$ are not 
manifest, but it simplifies considerably the computation of quantum 
corrections. 

One remark is in order here: due to the additional $SU(2)_R$ symmetry of the 
5D matter sector, Yukawa-type couplings between $\phi$ and $\psi$ or trilinear 
couplings between $\Phi$s are forbidden in the bulk. 


\subsection{The orbifolding and the low-energy spectrum}
If we want to recover the MSSM at low energy, we need  chiral zero modes for 
fermions.  To realize this, we will compactify the fifth dimension on the 
orbifold $S^1/Z_2$. The orbifold construction is crucial in order to obtain
chiral zero modes from a vector-like 5D theory.  

\begin{figure}[htb]
\begin{center}
  \mbox{\epsfxsize=0.8\textwidth\epsffile{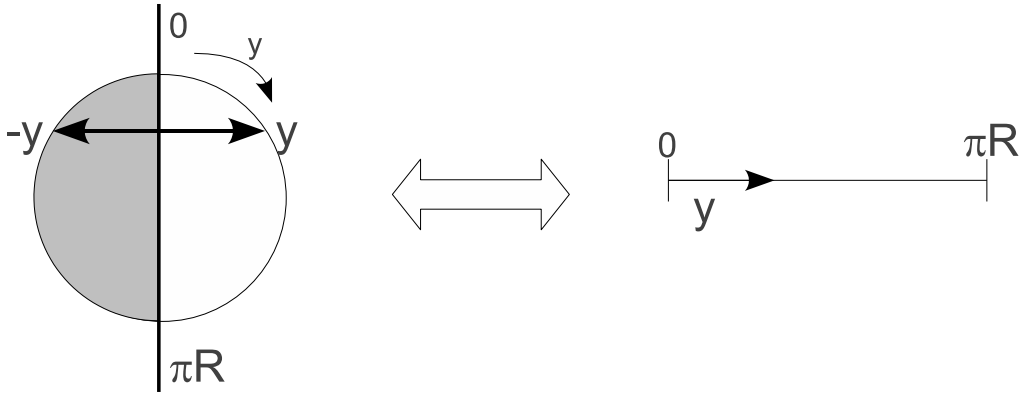}}
  \end{center}
\caption[]{\label{orbifold} The orbifold projection ($S^1/Z_2$). The physical 
interval is $0\leq y\leq\pi R$.}
\end{figure}

\begin{figure}[htb]
\begin{center}
  \mbox{\epsfxsize=0.5\textwidth\epsffile{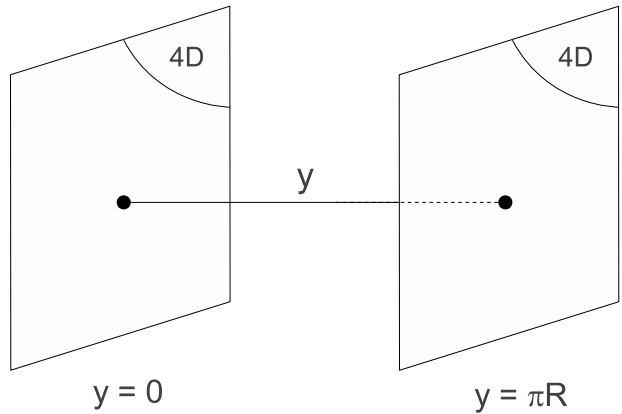}}
  \end{center}
\caption[]{\label{branes} Branes of the $S^1/Z_2$ orbifold.}
\end{figure}

The $Z_2$ symmetry identifies  
$y\to-y$, and reduces the physical interval to $[0,\pi R]$ where $R$ is 
the radius of the circle (see Fig.~\ref{orbifold}).
We have two orbifold fixed points
invariant under the $Z_2$
transformation, namely $y=0=-y$ and $y=\pi R=-\pi R+ 2\pi R=-y$. These 
fixed points, called branes (cf. Fig.~\ref{branes}), break the translational
invariance in the fifth dimension and therefore the momentum 
conservation along
$y$ and thus part of the 5D supersymmetry. We will choose the 
transformation properties of the fields and the interactions such that
4D Lorentz symmetry, the $Z_2$ orbifold symmetry  
and one 4D $\mathcal{N} = 1$ supersymmetry are preserved. 
The $\chi$-field should then be 
odd under the $Z_2$ symmetry because it appears together with a derivative
$\partial_y$, whereas $V$ is even. For the two matter superfields, we choose
$\Phi$ to be even and the conjugate $\Phi^c$ to be odd. This is a pure 
convention. Note that only the even fields have zero modes.
The Fourier decomposition 
of the fields reads:
\begin{eqnarray*}
V(x,y) &=& \frac{1}{\sqrt{\pi R}}\left[V^{(0)}(x)+\sqrt{2}\sum_{n\geq 1}V^{(n)}(x)\cos\left(\frac{ny}{R}\right)\right] \\
\chi(x,y) &=& \sqrt{\frac{2}{\pi R}}\sum_{n\geq 1}\chi^{(n)}(x)\sin\left(\frac{ny}{R}\right) \\
\Phi(x,y) &=& \frac{1}{\sqrt{\pi R}}\left[\Phi^{(0)}(x)+\sqrt{2}\sum_{n\geq 1}\Phi^{(n)}(x)\cos\left(\frac{ny}{R}\right)\right] \\
\Phi^c(x,y) &=& \sqrt{\frac{2}{\pi R}}\sum_{n\geq 1}\Phi^{c(n)}\sin\left(\frac{ny}{R}\right)~,
\end{eqnarray*}
where we normalised the massive KK states to have canonical kinetic terms. 
At energies well below the scale $R^{-1}$, where the massive Kaluza-Klein 
states decouple, only the zero modes remain in the spectrum and we assume
that physics is described by the usual MSSM. Thus, the matter superfields (and 
Higgs superfields) of the MSSM will be identified with a $\Phi^{(0)}$ superfield,
and the gauge fields with a $V^{(0)}$ mode. From the above decomposition
it becomes obvious that the orbifold prescription breaks the
original $\mathcal{N}=2$ supersymmetry which relates 
$\Phi$ to $\Phi^c$ and $V$ to $\chi$.

Compactifying on $S_1/Z_2$ we have to introduce two Higgs hypermultiplets
in the 4D $\mathcal{N}=2$ language if we want to have zero modes
corresponding to the two Higgs superfields of the MSSM, $H_u$ and $H_d$.  
Introducing an additional $Z_2'$ symmetry it is possible to obtain two
zero mode Higgs superfields starting from one Higgs hypermultiplet~\cite{BHN}.
For simplicity we will stay with only one $Z_2$ symmetry here.

\subsection{Flavour physics on the brane}

As stated above, Yukawa couplings in the bulk are forbidden by the 5D 
$\mathcal{N}=1$ supersymmetry. However, they can be introduced {\it 
on the branes}, which are 4D subspaces with reduced supersymmetry.
We will write the following interaction terms, called {\it brane interactions},
containing Yukawa-type couplings. 
\beq
S_{brane}=\int\mbox{d}^8z\mbox{d}y\delta(y)\left\{ \left(\frac{1}{6}
\tilde{\lambda}_{ijk}\Phi_i\Phi_j\Phi_k + \frac{1}{4}\tilde{\kappa}_{ij}L_iH_uL_jH_u\right)\delta(\bar{\theta}) 
+ \mbox{h.c.} \right\}~,
\label{Sbrane}
\eeq
where $\Phi_i$ represents a matter superfield, $L$ is the lepton doublet
superfield, and $H_u$ one Higgs superfield. The last term corresponds to the 
effective neutrino mass operator, with
dimensionful coupling $\tilde{\kappa}_{ij}$. 

Note that we do not write any interactions with conjugate superfields 
$\Phi^c$ on the brane which would be allowed by gauge interactions. But, as 
$\Phi^c$ vanishes on the branes, these interactions simply vanish. Moreover, 
we could have introduced independent interactions on the  
$\pi R$ brane. Below we will briefly comment on this possibility, but for our
numerical analysis we restrict ourselves to~(\ref{Sbrane}).

\section{Beta functions}
In this section we will derive the beta functions for the Yukawa
couplings and the coupling of the neutrino mass operator assuming that 
no other operators (generated in the evolution) affect this behaviour. We 
will begin with recalling briefly the 4D result, mainly
in order to set up notations and to explain the method.    
\subsection{Usual 4D result}
Due to the non-renormalisation
theorem~\cite{NRtheorem}, the beta functions for the couplings
of the operators in the superpotential are governed by the wave
function renormalisation constants $Z_{ij} = 1 + \delta Z_{ij}$. 
These relate the bare to
the renormalised superfields,
\beq
\Phi^{(i)}_B=\sum_{j=1}^{N_{\Phi}} Z_{ij}^{1/2}\Phi^{(j)}_R~.
\eeq
The sum runs here over all $N_\Phi$ chiral superfields of the model.

In a generic 4D super-Yang-Mills theory, the result for the wave
function renormalisation constant at one loop and in $d = 4-\epsilon$ dimension reads~\cite{Westbook}:
\beq
-\delta
Z_{ij}=\frac{1}{16\pi^2}\frac{2}{\epsilon}
\left(\sum_{k,l=1}^{N_{\Phi}}\frac{1}{2}\lambda^*_{ikl}\lambda_{jkl}-2\sum_{n=1}^{N_{g}}g_n^2C_2(R_n^{(i)})\delta_{ij}\right)~.
\label{Zmssm}
\eeq
We have written the interaction as
$\lambda_{ijk}\Phi^{(i)}\Phi^{(j)}\Phi^{(k)}$ and the sum over $n$ runs
over all gauge groups of the theory. The group-theoretical constants
$C_2(R)$ are defined as
\beq
C_2(R) \delta_{ab} = \sum_A (T^A T^A)_{ab}~,
\eeq
where $T^A$ are matrix representations of the generators of the
gauge group corresponding to the irreducible representation $R$ under
which the fields $\Phi_i$ transform. 
The result at two-loops can be found in~\cite{West}.

For future reference we recall here the result at 
one-loop for the the beta function of $\kappa'$:
\beq
(16\pi^2)\beta_{\kappa}=
\left(6\mathrm{Tr}(Y_u^{\dag}Y_u)-\frac{6}{5}g_1^2-6g_2^2\right)\kappa' +
\left[Y_e^{T}Y_e^*\right]\kappa' + \kappa'\left[Y_e^{\dag}Y_e\right]\\
\label{kappageneral}
\eeq
The two-loop result can be found 
in Ref.~\cite{AntuschRatz}. The expression for the Yukawa couplings
can be derived analogously.

\subsection{5D result}
We will now derive the results in the case of the five-dimensional
MSSM discussed above. 
To deal with the
running in extra-dimensional theories, intrinsically
non-renormalisable, we briefly remind the point of view introduced in
Ref.~\cite{DDG}. The theory is treated as a chain of effective field
theories where we decouple all the excitations whose mass exceeds the
energy $\mu$ we are interested in. Hence the greater the energy, the larger
the number of states considered, which creates to a very good
approximation a power law running of the couplings (under the
condition that the energy scale of interest is at least about an order of 
magnitude
larger than $R^{-1}$). See appendix \ref{integrals} for more details.

Higgs superfields and gauge superfields will always
propagate into the fifth dimension. Different possibilities of
localisation for the matter superfields will be studied by taking the two
limiting cases of superfields containing the SM fermions in the bulk or all
superfields containing SM fermions restricted to the brane, respectively.
We will begin with the case where all matter fields propagate in the
bulk.
\subsubsection{All matter superfields propagate in the bulk}
\label{secbulk}
If all matter chiral superfields of the MSSM are allowed to propagate
in the fifth dimension, we find the wave function renormalisation
constant of a matter (chiral) superfield:
\beq
-(16\pi^2)\,\delta Z^{5D}_{\Phi}=
\left(-8\sum_{n=1}^{N_{g}}g_n^2C_2(R_n^{(i)})\delta_{ij}\right)\Lambda R
+\left(2\pi\sum_{k,l=1}^{N_{\Phi}}\lambda^*_{ikl}\lambda_{jkl}\right)\Lambda^2R^2~.
\label{Zsusy5D}
\eeq 
Here $R$ corresponds to the radius of the compactified fifth
dimension and $\Lambda$ is a cutoff parameter. We only retained the
contributions which diverge in the limit $\Lambda \to \infty$, see
appendix~\ref{integrals}, where we discuss the evaluation of the sums
over KK states.  The same result is
obtained for the Higgs superfield.  A collection of explicit
expressions for the different wave function renormalisation constants
can be found in appendix~\ref{zbulk}.

As in the previous section, the beta functions can be directly
calculated from the above expression for the wave function
renormalisation constants. Within the model discussed in
Section~\ref{model} we obtain for the beta function of $\kappa = \tilde\kappa/(\pi R)^2$ at one loop:
\beqn
(16\pi^2)\beta_{\kappa}&=&\left((-\frac{12}{5}g_1^2-12g_2^2)\Lambda R +24\pi
\mathrm{Tr}(Y_u^{\dag}.Y_u)\Lambda^2R^2       
\right)\kappa \nonumber\\
&+&
\left(\left[Y_e^{T}Y_e^*\right]\kappa +
\kappa\left[Y_e^{\dag}Y_e\right]\right) 4 \pi \Lambda^2 R^2~.
\label{betakappa5D1}
\eeqn 
The beta functions for Yukawa couplings are given by:
\beqn
(16\pi^2)\beta_{Y_d}&=&Y_d(3\mathrm{Tr}(Y_d^{\dag}Y_d)+\mathrm{Tr}(Y_e^{\dag}Y_e)+3Y_d^{\dag}Y_d+Y_u^{\dag}Y_u)4\pi\Lambda^2R^2
\nonumber\\
&-&Y_d\left(\frac{14}{15}g_1^2+6g_2^2+\frac{32}{3}g_3^2\right)\Lambda
R \\
(16\pi^2)\beta_{Y_u}&=&Y_u(3\mathrm{Tr}(Y_u^{\dag}Y_u)+3Y_u^{\dag}Y_u+Y_d^{\dag}Y_d)4\pi\Lambda^2R^2
\nonumber\\
&-&Y_u\left(\frac{26}{15}g_1^2+6g_2^2+\frac{32}{3}g_3^2\right)\Lambda
R \\
(16\pi^2)\beta_{Y_e}&=&Y_e(3\mathrm{Tr}(Y_d^{\dag}Y_d)+\mathrm{Tr}(Y_e^{\dag}Y_e)+3Y_e^{\dag}Y_e)4\pi\Lambda^2R^2
\nonumber\\ &-&Y_e\left(\frac{18}{5}g_1^2+6g_2^2\right)\Lambda R
\label{beta5D1}
\eeqn
Note that all beta functions contain a term quadratic in $\Lambda
R$. This term will dominate the evolution of the Yukawa couplings and
of $\kappa$. They will evolve much more rapidly than the gauge
couplings, which only contain a linear term. This linear term arises
from the sum over the tower of KK states. Since the Yukawa
interactions and the effective neutrino mass operator are localised on
the brane, we have to sum over two towers of KK excitations giving
rise to the quadratic term. This effect has already been noticed in
Ref.~\cite{BHN}, where limitations of the model due to the
quadratic running have been mentioned, too. For the same reason at higher orders 
in the loop expansion higher powers of $\Lambda R$ appear which limit the validity 
of the present approach. The top
Yukawa coupling can become non-perturbative before the
gauge couplings and at rather low energy thus limiting 
considerably the range of validity of the present model. This will
become evident from the discussion of the numerical results in
Section~\ref{numerics}. We could have introduced another independent
interaction on the $\pi R$ brane. This would not change the general problem 
since it is not possible to mutually compensate the quadratic terms. Thus,
without allowing for Yukawa interactions in the bulk, which would break
the supersymmetry, it is not possible to avoid this quadratic running if there
are matter fields in the bulk. 

\subsubsection{Matter superfields on the brane}
\label{secbrane}
In this section we will discuss the results for the beta function in the case 
where all matter superfields are constrained to live on the 4D brane at 
$y = 0$, i.e. there are no KK excitations for the matter superfields. We thus
expect that the quadratic evolution due to the sum over two KK towers
will become milder. The renormalisation 
constant of a matter (chiral) superfield becomes:
\beq
-(16\pi^2)\,\delta Z^f_{\Phi}=
\left(-16\sum_{n=1}^{N_{g}}g_n^2C_2(R_n^{(i)})\delta_{ij}
+4\sum_{k,l=1}^{N_{\Phi}}\lambda^*_{ikl}\lambda_{jkl}\right)\Lambda R
\label{Zsusy5D2}
\eeq
Since the Higgs-type superfields can propagate in the bulk, in contrast to the 
matter superfields, the wave function renormalisation has no longer the same
structure. For example, the contribution containing the vector superfields
contains a sum over KK states for the Higgs superfield, whereas only the zero
mode enters in the case of the matter superfields. For the Higgs superfield we 
obtain
\beq
-(16\pi^2)\,\delta Z_{H}=
\left(-8\sum_{n=1}^{N_{g}}g_n^2C_2(R_n^{(i)})\delta_{ij}\right)\Lambda R
+\left(\sum_{k,l=1}^{N_{\Phi}}\lambda^*_{ikl}\lambda_{jkl}\right)\log\Lambda R
\label{Zsusy5D2higgs}
\eeq
Again, explicit expressions for the wave function renormalisation constants can be found in appendix~\ref{zbrane}. 

As in the two preceeding sections, we can derive the beta functions from the 
above equations for the wave function renormalisation constant. 
The beta function of $\kappa$ is given by
\beq
(16\pi^2)\beta_\kappa=\left((-\frac{18}{5}g_1^2-18g_2^2)\Lambda
R+6\mathrm{Tr}(Y_u^{\dag}.Y_u)\right) \kappa 
+ \left(\kappa Y_e^{\dag}.Y_e
+ Y_e^{T}.Y_e^* \kappa \right)4 \,\Lambda R~.
\label{BranebetaKappa5D}
\eeq
The beta functions for Yukawa couplings read
\beqn
(16\pi^2)\beta_{Y_d}&=&Y_d(3\mathrm{Tr}(Y_d^{\dag}Y_d)+\mathrm{Tr}(Y_e^{\dag}Y_e))
\nonumber\\
&+&Y_d\left(-\frac{19}{15}g_1^2-9g_2^2-\frac{64}{3}g_3^2+12Y_d^{\dag}Y_d+4Y_u^{\dag}Y_u\right)\Lambda R \\
(16\pi^2)\beta_{Y_u}&=&3Y_u\mathrm{Tr}(Y_u^{\dag}Y_u)
\nonumber\\
&+&Y_u\left(-\frac{43}{15}g_1^2-9g_2^2-\frac{64}{3}g_3^2+12Y_u^{\dag}Y_u+4Y_d^{\dag}Y_d\right)\Lambda R \\
(16\pi^2)\beta_{Y_e}&=&Y_e(3\mathrm{Tr}(Y_d^{\dag}Y_d)+\mathrm{Tr}(Y_e^{\dag}Y_e))
\nonumber\\ &+&Y_e\left(-\frac{33}{5}g_1^2-9g_2^2+12Y_e^{\dag}Y_e\right)\Lambda R 
\eeqn
As can be seen from the above equations neither $\kappa$ nor the Yukawa couplings contain a term quadratic in 
$\Lambda R$ any more, since there are 
no KK excitations for the matter superfields. This means that the range of validity of this
model will be wider since the couplings, in particular the top Yukawa
coupling, will not become non-perturbative at low energies (a few $R^{-1})$. This will
be discussed in more detail in the next section where numerical results will 
be presented. 
%
%
\section{Numerical results}
\label{numerics}
%
%
Within this section we will apply the beta functions derived in the
previous section to study the evolution of the different couplings. We
are particularly interested in the evolution of the neutrino mass
parameters. To that end we employ the MATHEMATICA package
REAP~\cite{Antuschetal} in a slightly modified version including our
models.  The SUSY threshold is taken at 1 TeV. Throughout the
discussion we will identify the cutoff $\Lambda$ with the energy scale
$\mu$ for the evolution.

Before coming to a detailed discussion of the numerical results, let
us mention that the discussion of the fixed point structure for the
mixing angles follows closely the MSSM one
\cite{Chankowski:1999xc,Casas:2003kh}. The evolution equations (given
in detail in Appendix \ref{conventions}) have the same structure as in
MSSM. We will come back to
this point in section~\ref{num:brane}. 

\subsection{Matter superfields in the bulk}
Let us begin with the model described in Sec.~\ref{secbulk}. The
corresponding beta functions have been given in
Eqs.~(\ref{betakappa5D1}, \ref{beta5D1}). In this case, as already
anticipated, the evolution is dominated by the quadratic terms. This
puts stringent limits on the model.  As soon as the energy scale
$\mu$ becomes of the order of $R^{-1}$, i.e. the couplings start to feel the
effect of the extra dimension, $y_t$ starts to increase rapidly and it
diverges for $\mu R\sim 2-3$. Moreover cubic and higher divergences arise
at higher orders in the loop expansion which can strongly affect the results.
This is due to the presence of sums on the KK excitations in the loops 
which are not restricted by any conservation of KK numbers at the vertices.

\begin{figure}[htb!]
\begin{center}
  \mbox{\epsfxsize=0.5\textwidth\epsffile{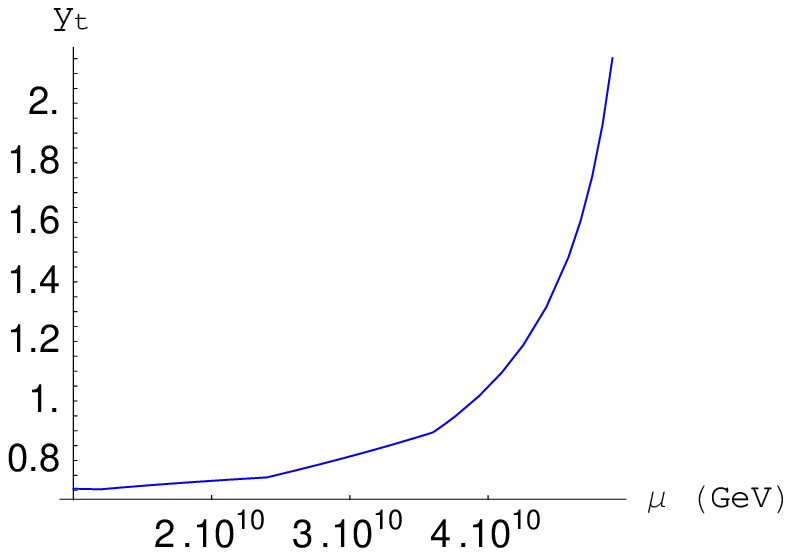} \epsfxsize=0.5\textwidth\epsffile{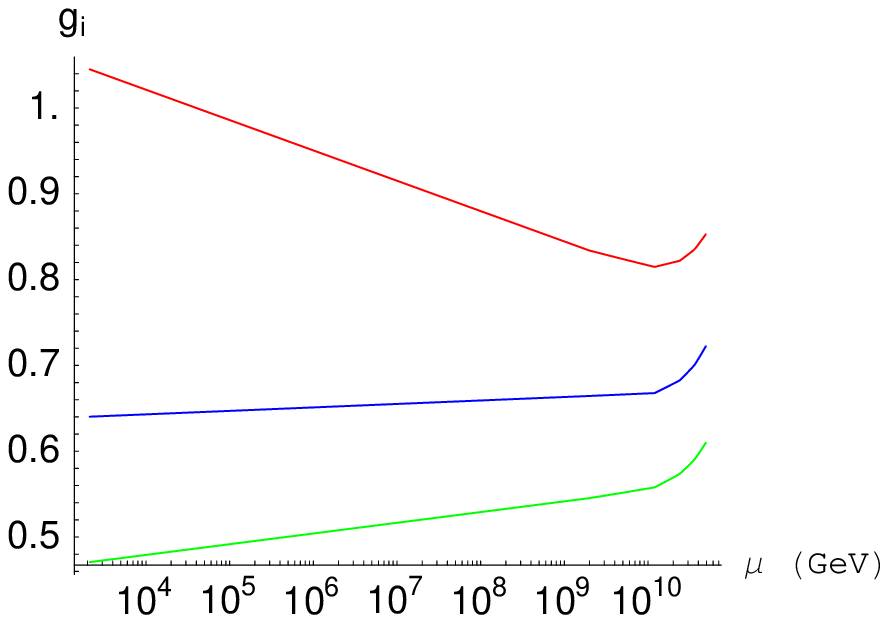}}
  \end{center}
\caption[]{\label{bulk} Evolution of the top Yukawa coupling (left
  panel) and the gauge couplings (right panel) $g_1$,
  $g_2$ and $g_3$ (from top to bottom) in 
the case $R^{-1}\sim 10^{10}$GeV and matter fields in the bulk. We clearly see  
that $y_t$ diverges before any perturbative unification is possible.}
\end{figure}

Throughout the calculation of the
beta functions, we have assumed that $\mu R \gg 1$. How reliable
is this approximation if we reach the perturbative limit of the model
already at $\mu R \sim 2-3$?  We checked the validity of the
approximation by integrating the beta functions without using the
power law approximation. This can be done numerically, adding at
every threshold the contributions from the new degrees of freedom. The
results do not change considerably.  For instance, $y_t$ diverges for
a slightly larger value of $\mu R\sim$4-5 (see Fig.~\ref{bulk}). 

However, a unification of the gauge couplings before the
non-perturbativity of the top Yukawa coupling is possible at a scale
$R^{-1}\geq 2.10^{14}$ GeV.  This scale is of the same order as the
standard 4D unification scale, such that the extra dimensional
scenario looses much of its interest. In addition, the rapid
divergence of the top Yukawa coupling prevents us from making any
reliable prediction on the effects of the extra dimension on the
neutrino sector.  We will thus not show any results for the neutrino
parameters within this model.

\subsection{Matter superfields on the brane}
\label{num:brane}
If we restrict the matter fields to the brane, we no longer face this problem. 
First of all, there is no divergent quadratic term at
1-loop. In addition, thanks to the large negative contribution of the
gauge couplings to $\beta_{Y_u}$, $y_t$ decreases. This allows for a
perturbative unification of the gauge sector at a value $\mu\sim 40
R^{-1}=M_{GUT}$ for $R^{-1}=10^3$ TeV.  We have also checked
explicitely that 2-loop terms are at most quadratic in the cut-off.
These terms are further suppressed by an extra factor of $16\pi^2$ so
that the results are presumably not modified within the range of
validity of the effective theory established at 1-loop.

This model is thus much more promising and worthwhile studying the
evolution of the neutrino mass parameters. We discuss our conventions
for the mixing matrix and masses together with the explicit
renormalisation group equations (RGE) in appendix~\ref{conventions}.

Here a few general comments are in order. The RGE for the neutrino
masses and mixings are similar to the 4D case since the beta functions
have a similar structure. Consequently the relevant parameters will be
essentially the same as in the 4D case. $\tan\beta$ plays an important
role as all the mixing angles and phases depend on $y_\tau$. In
addition, $m_1$ is important since the running of $\theta_{12}$ will
be stronger if $m_1 \sim m_2$. This is the reason why we generally
choose $\tan\beta\sim 50$ and $m_1\sim 0.1$~eV to explore the effects
of the extra-dimension.

\begin{figure}[htb!]
\begin{center}
  \mbox{\epsfxsize=0.6\textwidth\epsffile{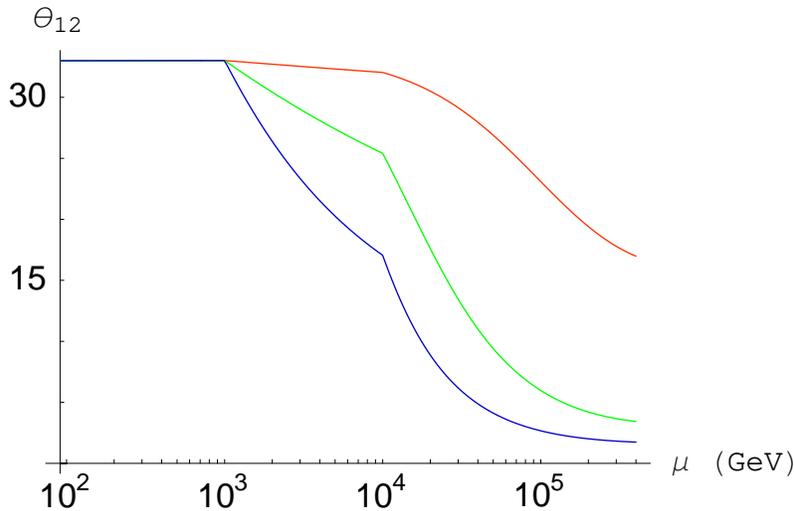}}
  \end{center}
\caption[]{\label{theta12} Running of $\theta_{12}$ for the values 
$\tan\beta=\{10,30,50\}$ (from top to bottom), $R^{-1}=10^4$~GeV, $m_1=0.1$~eV, $\theta_{13}=0$ 
and all phases vanish at $M_Z$.}
\end{figure}

In Fig.~\ref{theta12} we show the influence of
$\tan\beta$ on the evolution of $\theta_{12}$, all other parameters
are kept fixed. Depending on the value of $\tan\beta$, $\theta_{12}$ can
assume almost any value at the scale
where gauge coupling unification is achieved within our model. It
indicates the scale where new physics will come into play.
This sensitivity of $\theta_{12}$ to $\tan\beta$ has already been
noticed in 4D (cf for instance Ref.~\cite{Antusch:2003kp}). 
\subsubsection{Masses}
\begin{figure}[htb!]
\begin{center}
  \mbox{\epsfxsize=1.0\textwidth\epsffile{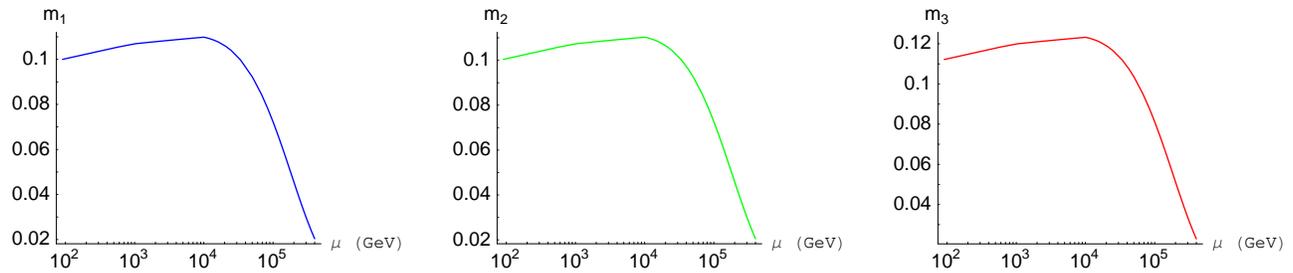}}
  \end{center}
\caption[]{\label{masses} Running of the three masses for the values 
$R^{-1}=10^4$~GeV, $m_1=0.1$~eV, $\theta_{13}=0$ and all phases vanish
at $M_Z$.}
\end{figure}

Still, the 5D case is intrinsically different from the 4D one and has
some particularities that are quite independent of any choice of low
energy parameters.  This is the case for the evolution of the masses
(Fig.~\ref{masses}). As can be seen in Fig.~\ref{masses} the
evolution has exactly the same form for the three masses and is much
sharper above $\mu \sim R^{-1}$, i.e. where the masses start to
feel the effect of the fifth dimension. This leads to a
reduction of up to a factor of the order of five for the masses in the
UV with respect to the values at low energies. This prediction is
extremely stable and can be explained as follows.  

The evolution of
the masses is governed by the following equation 
\beq \frac{d
m_i}{d(t=\ln\mu R)}=\dot{m}_i=\frac{1}{16\pi^2}\left[ \alpha +
4\mu Ry^2_\tau a_i \right]m_i~.
\eeq 
where the parameters $a_i$ induce a priori a non-universal behaviour
and the parameter $\alpha$ is detailed in appendix \ref{conventions},
in 4D and 5D. In contrast to the MSSM, the evolution in our case is
completely dominated by the universal part. The essential point is
that in the MSSM the positive contribution to $\alpha$, approximately
$6y_t^2$, is of the same order as the negative contribution from the
gauge part, leaving an $\mathcal{O}(1)$ factor. In the setup we are
interested in, however, the situation is completely different: on the
one hand $y_t$ decreases very rapidly and on the other hand the
contribution from the gauge part to $\alpha$ is multiplied by $12\mu
R$ with respect to the MSSM which makes it completely dominant
compared to any other contribution.  We can therefore write:
\beq 
\dot{m}_i \sim
\frac{1}{4\pi^2}\mu R\left(-\frac{18}{5}g_1^2-18g_2^2\right)m_i~.
\eeq 
This equation is universal. From this approximation we immediately see
that all masses decrease with increasing energy and eventually become
zero. 

\begin{figure}[htb!]
\begin{center}
  \mbox{\epsfxsize=1.0\textwidth\epsffile{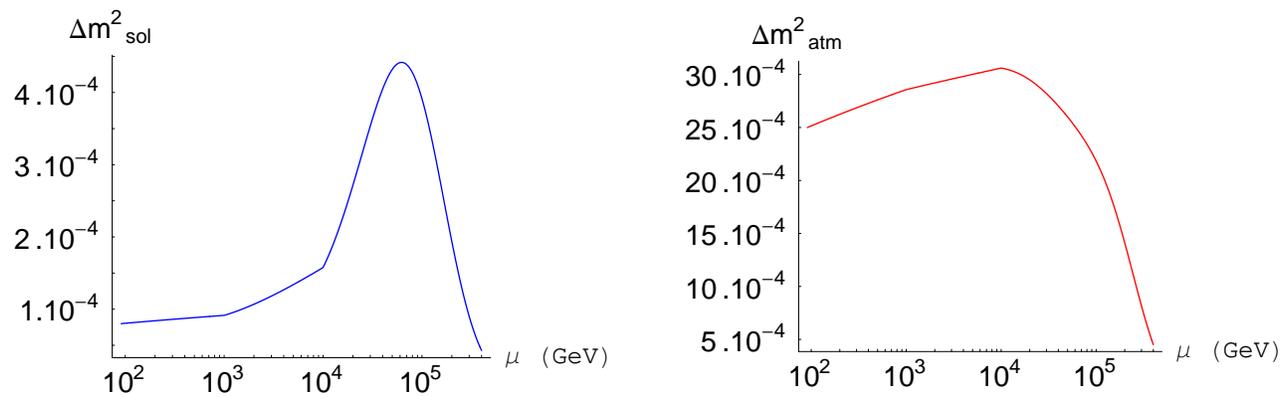}}
  \end{center}
\caption[]{\label{deltas} Running of $\Delta m_{sol}^2$ and 
$\Delta m_{atm}^2$ for the values $\tan\beta=50$, $R^{-1}=10^4$GeV, 
$m_1=0.1$eV and all phases vanish at $M_Z$.}
\end{figure}

This discussion can be extended to $\Delta m^2$, too, except in the
case where $m_1$ is not small compared with $m_2$ and $\tan\beta$ is
large. In the latter case we can check from the analytic formula,
\beq
\frac{d}{dt}\Delta m_{sol}^2=\frac{1}{8\pi^2}\left[ \alpha\Delta
m_{sol}^2 + 4\mu
Ry_\tau^2\left[2s_{23}^2(m_2^2c_{12}^2-m_1^2s_{12}^2) +
\mathcal{O}(\theta_{13})\right] \right] 
\eeq 
that for $\sqrt{\Delta m_{sol}^2}\ll m_1$ the first term does not
necessarily dominate. Still, because all the masses decrease rapidly,
$\Delta m_{sol}^2$ has to stop growing at some point. This is indeed
the case, see Fig.~\ref{deltas}.
\subsubsection{Mixing angles}
Note also that, assuming a hierarchical spectrum ($m_1=0.001$eV) and
$\tan\beta=10$, $\theta_{12}$ will vary twice more than in the MSSM
between $M_Z$ and $M_{GUT}$, provided $\theta_{13}$ is rather large
($\simeq 9$ degrees).

\begin{figure}[htb]
\begin{center}
  \mbox{\epsfxsize=0.6\textwidth\epsffile{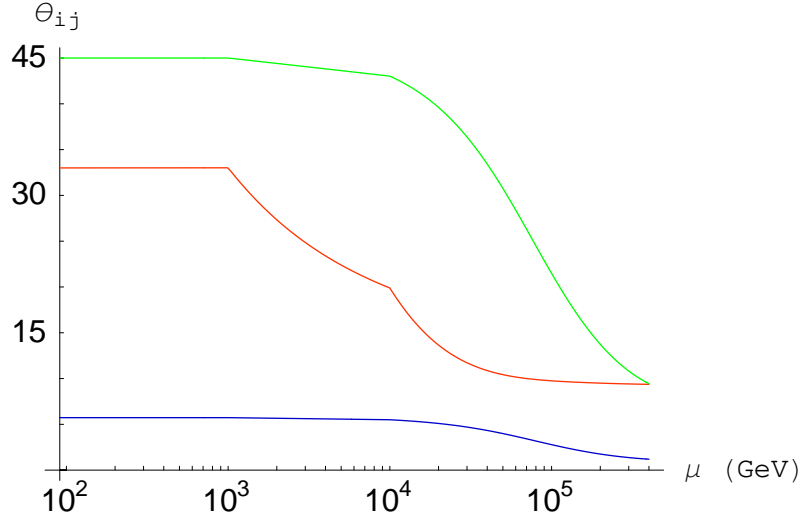}}
  \end{center}
\caption[]{\label{thetaij} Running of $\theta_{12}$ (red/grey), 
$\theta_{13}$ (blue/dark grey) and $\theta_{23}$ (green/light grey) for 
the values $\tan\beta=50$, $R^{-1}=10^4$GeV, $m_1=0.1$eV and no phases at 
$M_Z$}
\end{figure}

Another interesting observation is that $\theta_{23}$ can become much
smaller than in the 4D case. This allows for a unification of
$\theta_{12}$ and $\theta_{23}$ at a value below ten degrees.  This
can be seen in Fig.~\ref{thetaij} where we display the three mixing
angles for $\tan\beta = 50$ and $m_1 = 0.1$eV.  

\begin{figure}[htb!]
\begin{center}
  \mbox{\epsfxsize=1\textwidth\epsffile{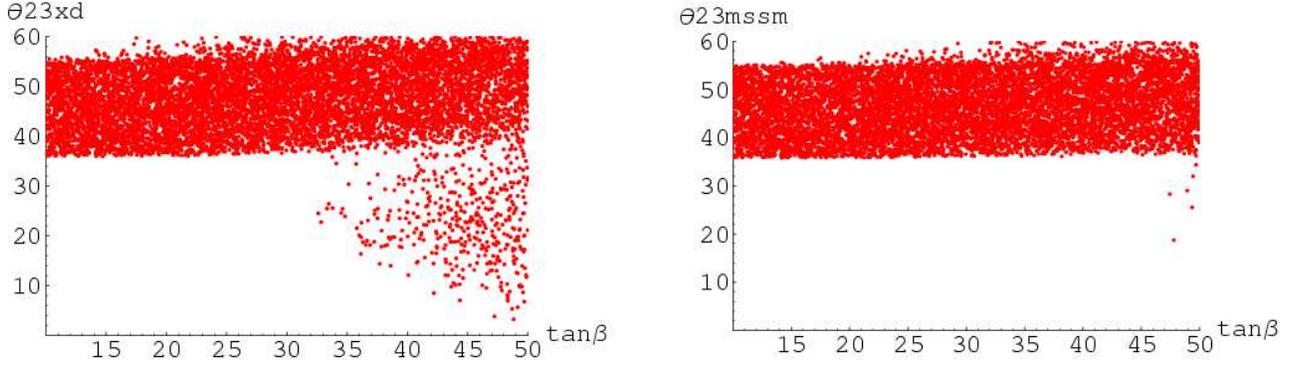}}
  \end{center}
\caption[]{\label{scatterplot2} 
Comparison of $\theta_{23}$ at the cut-off scale as a function of $\tan
 \beta$ in our 5D model and in 4D MSSM for random phases and
 $0.01<m_1<0.1$ eV with inverse hierarchy. For smaller values of $m_1$
 the spread is reduced. 
 }
\end{figure}

\begin{figure}[htb!]
\begin{center}
  \mbox{\epsfxsize=1\textwidth\epsffile{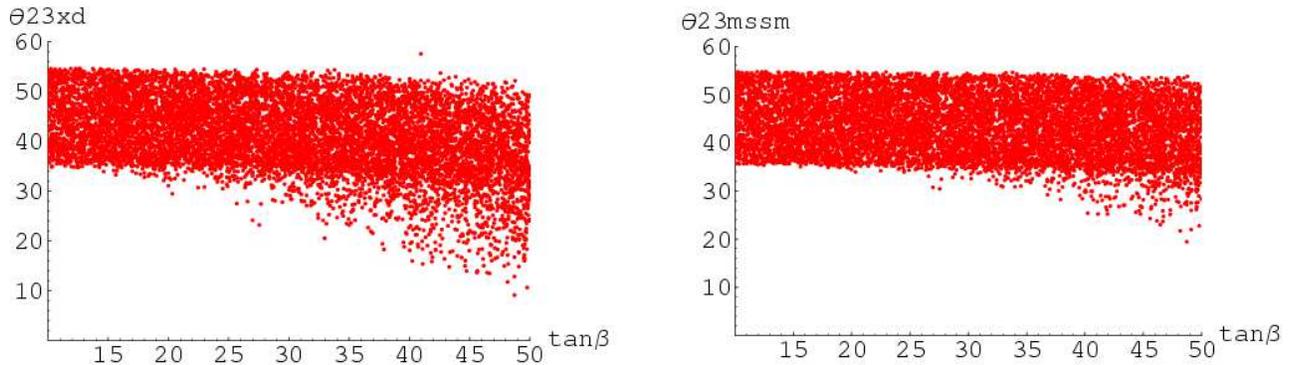}}
  \end{center}
\caption[]{\label{scatterplot3} 
Same as Fig.~\ref{scatterplot2} for normal hierarchy.
 }
\end{figure}

\begin{figure}[htb!]
\begin{center}
  \mbox{\epsfxsize=1\textwidth\epsffile{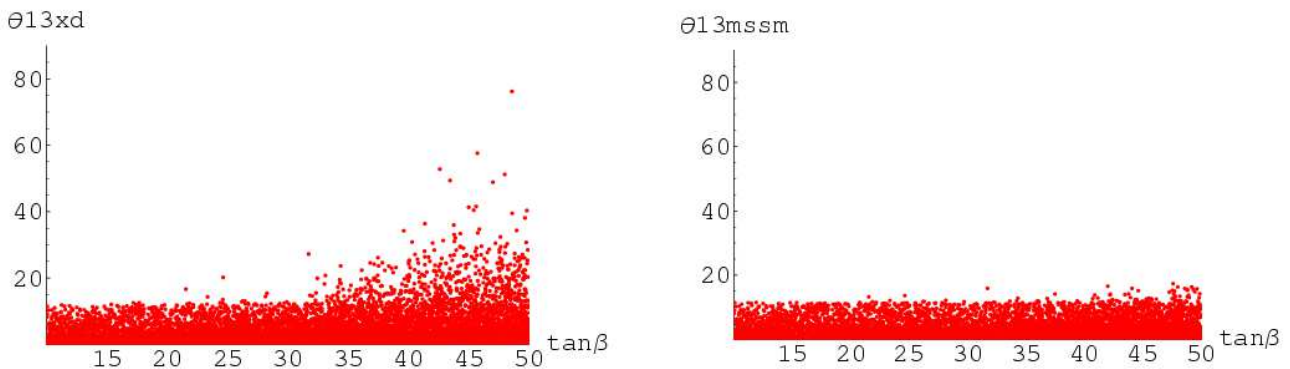}}
  \end{center}
\caption[]{\label{scatterplot1} 
 Same as Fig.~\ref{scatterplot2} for $\theta_{13}$.
 }
\end{figure}

In Figs.~\ref{scatterplot2} and \ref{scatterplot3} we show results for
$\theta_{23}$ at the cut-off scale in our 5D model and in the 4D MSSM.
Fig.~\ref{scatterplot2} corresponds to an inverse hierarchy whereas
Fig.~\ref{scatterplot3} corresponds to normal hierachy.  For the
results shown we have $0.01 \mathrm{eV}<m_1<0.1 $eV. For smaller
masses the spread is reduced. In general, compared with the MSSM we 
obtain new interesting points where $\theta_{23}$ is small, ($\leq$ 20 degrees)
for quite large $\tan\beta\sim 30-50$ and $m_1\sim 0.1$eV.  Note that it is even easier to
get such small $\theta_{23}$ with an inverted hierarchy. In this case
we can also find a portion of the parameter space where $\theta_{13}$
can be rather large, see Fig.~\ref{scatterplot1}. This is in sharp
contrast with the 4D MSSM, thus suggesting the possibility that
$\theta_{13}\simeq\theta_{12}$ at the cut-off scale with values around roughly
30 degrees. There are even some points with larger values, up to 45
degrees.

Let us note that, although previous studies \cite{Agarwalla:2006dj} have
investigated the possibility of having CKM-like values for the PMNS
angles at high energies, they required even larger values
$0.1\mbox{eV}<m_1<0.6\mbox{eV}$ creating some tension with
cosmological bounds. Here a value $m_1\leq0.1$eV is sufficient, which 
is new compared with the 4D case. 
This opens the interesting possibility to find a set of
parameters which generate a maximal mixing from a small
CKM-like mixing at some new physics scale or even 
allow for unification of the three mixing angles at high energy. 


Despite these interesting features we note that the
region in parameter space most studied here is the region where
$m_1\simeq m_2$. In this case the low energy values of the mixing
angles (in particular $\theta_{12}$) are very sensitive to the precise
value of $m_1$ at the high energy scale (at the percent level). 
\par\hfill\par
As mentioned above, we have in principle the same fixed points for the
angles as in the 4D MSSM. In the case where $\theta_{13} = 0 $ it can
be seen immediately from the equations that $\theta_{23} = \theta_{12}
= 0$ are fixed points. For nonzero $\theta_{13}$ the fixed points are
less obvious, but as the structure of the equations does not change
with respect to the MSSM, the result is in principle the same. It leads
to fixed points predicting the
relation~\cite{Chankowski:1999xc,Casas:2003kh}
\begin{equation}
\sin^22\theta_{12} = \frac{s^2_{13} \sin^2 2 \theta_{23}}{(s_{23}^2
  c_{13}^2 + s_{13}^2)^2} \quad \mathrm{with}\quad s_{23}^2 =
  \frac{1}{2} \Big(1 \pm \sqrt{1 - \sin^2 2 \theta_{23}}\Big)
\end{equation}
for different patterns for the coefficients.  The coefficients change with
respect to the 4D MSSM and therefore the way we can approach the fixed
points. Two essential differences between the 5D and 4D MSSM case
should be noted in that context.  First, the coefficient $C$ governing
the evolution, see appendix~\ref{conventions}, is $C~=~1$ in the 4D
MSSM whereas it is $4
\mu R$ in the 5D case. Second, the numerical coefficients entering the
equations for the angles depend on the masses and become almost
constant if we enter the universal regime for the running of the
masses discussed above. This can be understood as follows. Neglecting
the phases, these coefficients can be written as $(m_i+m_j)/(m_i -
m_j)$ such that the indirect dependence on $\mu$ via the masses
cancels out within the universal regime. From our numerical analysis
we find that in general the fixed points are not reached within the
range of validity of the present model. Increasing $\tan\beta$
accelerates the evolution such that the angles come closer to the
fixed points at the cut-off scale. For very large values, $\tan\beta >
50$, we reach in some cases a fixed point.

\section{Conclusions}
The aim of this paper was to perform a study of the neutrino sector
in a simple supersymmetric extra-dimensional framework. We have presented
beta functions for Yukawa couplings and $\kappa$, the coupling of the
effective neutrino mass operator, within two distinct setups in
the context of a five-dimensional MSSM on $S^1/Z_2$. In the first case,
all fields associated with the SM fermions are allowed to propagate
in the fifth dimension whereas in the second they are restricted
to the brane.

Due to the 5D $\mathcal{N}=1$ supersymmetry, Yukawa interactions are 
forbidden in the bulk and
must be introduced on the branes. Within the first model, this induces
a quadratic running for Yukawa couplings and $\kappa$.  $y_t$ then
becomes non-perturbative already at rather low energies and even
before the gauge couplings. This strongly limits the range of validity
of the model.

Within the second scenario the dependence on the energy scale is only
linear and Yukawa couplings remain perturbative until gauge coupling
unification. The evolution of neutrino masses and mixings shows
interesting possibilities to explain the observed values at low
energies from some specific scenario at high energies. As a generic
prediction, neutrino masses are reduced up to a factor of five at the
unification scale with respect to their values at low energies.
From a top-down point of view, one can
radiatively generate a large mixing pattern at low energy starting
with small values of $\theta_{23}$ and $\theta_{12}$ at a high energy
scale with values for $m_1$ consistent with cosmological bounds. It is also 
possible to generate a small $\theta_{13}$ at $M_Z$ from large values at 
the cut-off scale.

\section*{Acknowledgements}
The authors are grateful to Paramita Dey and REAP authors for useful 
discussion and correspondences.
%
%
\appendix
%
%
\section{Conventions for neutrino masses and mixing parameters}
\label{conventions}
Within this section we would like to stress our conventions for the
mixing angles and phases and briefly discuss different scenarios for
neutrino masses. The mixing matrix which relates gauge and mass
eigenstates is defined to diagonalise the neutrino mass matrix in the
basis where the charged lepton mass matrix is diagonal. It is usually
parameterised as follows~\cite{Maki:1962mu}:
\begin{eqnarray}
U &=& \mathrm{diag}(e^{i\delta_e},e^{i \delta_\mu},e^{i \delta_\tau})\,
U_{\mathit{MNS}}~, \quad \mathrm{with} \nonumber\\
U_{\mathit{MNS}} &=& \left(\begin{array}{ccc} c_{12} c_{13} & s_{12}
    c_{13} & s_{13} e^{-i \delta} \\ -s_{12} c_{23} - c_{12} s_{23}
    s_{13} e^{i\delta} & c_{12} c_{23} - s_{12} s_{23} s_{13} e^{i
      \delta} & s_{23} c_{13} \\ s_{12} s_{23} - c_{12} c_{23} s_{13}
    e^{i \delta} & - c_{12} s_{23} - s_{12} c_{23} s_{13} e^{i\delta}
    & c_{23} c_{13} \end{array}\right) \mathrm{diag}(e^{-i
  \phi_1/2},e^{-i \phi_2/2},1)~.\nonumber \\
\end{eqnarray}
where $c_{ij} = \cos\theta_{ij}, s_{ij} = \sin\theta_{ij}$. We follow
the conventions of Ref.~\cite{Antusch:2003kp} to extract mixing
parameters from the $MNS$ matrix. 

Experimental information on neutrino mixing parameters and masses is
obtained mainly from oscillation experiments. The most simple
interpretation of these oscillation data is in terms of massive
neutrinos. There are essentially three types of experiments providing
us with data: solar neutrino experiments (Kamiokande and
Superkamiokande), looking for a deficit of $\nu_e$ from the sun,
atmospheric neutrino experiments looking for a deficit of $\nu_\mu$ and
$\bar\nu_\mu$ produced by cosmic rays in the earth's atmosphere and
reactor experiments looking for the neutrino flux from a
reactor. Present data
are compatible with oscillations between the three known neutrino
flavours. In general $\Delta m^2_{\mathit{atm}}$ is assigned to a mass
difference between $\nu_3$ and $\nu_2$ whereas $\Delta
m^2_{\mathit{sol}}$ to a mass difference between $\nu_2$ and $\nu_1$.
Current values~\cite{neutrinodata} are summarised in Table~\ref{tableexp}. 
Data indicate that $\Delta m^2_{\mathit{sol}} \ll \Delta
m^2_{\mathit{atm}}$, but the masses themselves are not
determined. Either they follow the hierarchical scenario with $ \Delta
m^2_{\mathit{atm}} \approx m^2_{\nu_3} \gg m^2_{\nu_{2(1)}} \gg
m^2_{\nu_{1(2)}}$ or (partly) degenerate scenarios with masses
approximately equal.\\
\begin{center}
\begin{table}[h!]
\begin{tabular}{c|c}
Parameter & Value (90\% CL) \\ \hline
$\sin^2(2\theta_{12})$ & $0.86(^{+0.03}_{-0.04})$ \\
$\sin^2(2\theta_{23})$ & $>0.92$ \\
$\sin^2(2\theta_{13})$ & $<0.19$ \\
$\Delta m^2_{\mathit{sol}}$ & $(8.0^{+0.4}_{-0.3})\times 10^{-5}eV^2$ \\
$\Delta m^2_{\mathit{atm}}$ & 1.9 to 3.0$\times 10^{-3}$ $eV^2$ \\
\end{tabular}
\label{tableexp}
\end{table}
\end{center}

With these conventions we can derive the approximate equations for the mixing angles and masses. The derivation is similar to the one in ref. \cite{Antusch:2003kp}, and with the same notations we write the result~:

\begin{eqnarray*}
\dot{m}_1 &=& \frac{1}{16\pi^2}[\alpha + Cy^2_\tau(2s_{12}^2s_{23}^2 + \mathcal{O}(\theta_{13}))]m_1\\
\dot{m}_2 &=& \frac{1}{16\pi^2}[\alpha + Cy^2_\tau(2c_{12}^2s_{23}^2 + \mathcal{O}(\theta_{13}))]m_2\\
\dot{m}_3 &=& \frac{1}{16\pi^2}[\alpha + Cy^2_\tau\times 2c_{13}^2c_{23}^2]m_3\\
\dot{\theta}_{12} &=& -\frac{Cy_\tau^2}{32\pi^2}\sin(2\theta_{12})s_{23}^2\frac{|m_1e^{i\phi_1}+m_2e^{i\phi_2}|^2}{\Delta m^2_{sol}} + \mathcal{O}(\theta_{13})\\
\dot{\theta}_{13} &=& \frac{Cy_\tau^2}{32\pi^2}\sin(2\theta_{12})\sin(2\theta_{23})\frac{m_3}{\Delta m^2_{atm}(1+\zeta)}\\
&& \qquad \times[m_1\cos(\phi_1-\delta) - (1+\zeta)m_2\cos(\phi_2-\delta)-\zeta m_3\cos\delta] + \mathcal{O}(\theta_{13})\\
\dot{\theta}_{23} &=& -\frac{Cy_\tau^2}{32\pi^2}\sin(2\theta_{23})\frac{1}{\Delta m^2_{atm}}\left[c_{12}^2|m_2e^{i\phi_2}+m_3|^2 + s_{12}^2\frac{|m_1e^{i\phi_1}+m_3|^2}{1+\zeta}\right] + \mathcal{O}(\theta_{13})
\end{eqnarray*}
where $\Delta m^2_{sol}=m_2^2-m^2_1$, $\Delta m^2_{atm}=m_3^2-m^2_2$, and $\zeta=\Delta m^2_{sol}/\Delta m^2_{atm}$.
The main difference here lies in the expressions of $\alpha$ and $C$, which are in the model with matter superfields on the brane:

\beq
\alpha=\left[\left(-\frac{18}{5}g_1^2-18g_2^2\right)\mu R + 6\mbox{Tr}(Y_u^\dagger.Y_u)\right] \qquad\qquad C=C(\mu)=4\mu R
\eeq

\noindent This has to be compared with the MSSM coefficients:

\beq
\alpha=\left(6\mbox{Tr}(Y_u^\dagger.Y_u)-\frac{6}{5}g_1^2-6g_2^2\right) \qquad\qquad C=1
\eeq

\section{Complete Lagrangian and Feynman rules}

Here we write the complete 5D action of the model where all fields can
propagate in the bulk~\cite{Flacke}:

\begin{eqnarray*}
S_{gauge} &=& \frac{\mathrm{Tr}}{C_2(G)}\int \mbox{d}^8z\mbox{d}y\left\{ -V\Box(P_T-\frac{1}{\xi}(P_1+P_2))V - V\partial_5^2 V - \frac{\xi}{2}\bar{\chi}\frac{\partial_5^2}{\Box}\chi + \frac{1}{2}\bar{\chi}\chi \right. \\ 
& & + \frac{\tilde{g}}{4}(\bar{D}^2D^\alpha V)[V,D_\alpha V] + \tilde{g}\left(\partial_5 V[V,\chi+\bar{\chi}]-(\chi+\bar{\chi})[V,\chi+\bar{\chi}]\right) \\
& &+\left. \mathcal{O}(g^2) +ghosts\phantom{\frac{1}{1}}\!\!\!\right\} \\
\\
S_{matter} &=& \int\mbox{d}^8z\mbox{d}y\left\{ \bar{\Phi}_i\Phi_i +
\Phi^c_i\bar{\Phi}^c_i + \Phi^c_i\partial_5\Phi_i \delta(\bar{\theta}) -
\bar{\Phi}_i\partial_5\overline{\Phi}_i^c\delta(\theta) \right. \\
& & \left. + \tilde{g}(2\bar{\Phi}_iV\Phi_i - 2\Phi_i^cV\bar{\Phi}^c_i + 
\Phi^c_i\chi \Phi_i\delta(\bar{\theta}) + \bar{\Phi}_i\bar{\chi}\bar{\Phi}^c_i\delta(\theta)) \right\} \\
\\
S_{brane} &=& \int\mbox{d}^8z\mbox{d}y\delta(y)\left\{ \left(\frac{1}{6}\tilde{\lambda}_{ijk}\Phi_i\Phi_j\Phi_k + \frac{1}{4}\tilde{\kappa}_{ij}L_iH_uL_jH_u\right)\delta(\bar{\theta}) + \mbox{h.c.} \right\}
\end{eqnarray*}
where we separated the pure gauge, the coupling of matter to gauge and
the Yukawa sector. The latter is localised on the brane in order to
respect 5D SUSY. $\xi$ is the
gauge fixing parameter.

In order to consider the 4D effective theory we compactify this action by 
expanding the fields as in section~\ref{model} and we keep only the terms 
that will be of interest to renormalise the Yukawa beta functions:

\begin{eqnarray*}
S_{gauge} &=& \int\mbox{d}^8z\left\{ -V_a^{(0)}\Box(P_T-\frac{1}{\xi}(P_1+P_2))V_a^{(0)} - \sum_{n\geq 1}V_a^{(n)}\Box(P_T-\frac{1}{\xi}(P_1+P_2))V_a^{(n)} \right. \\
& & + \left.\sum_{n\geq 1}\frac{n^2}{R^2}V_a^{(n)}V_a^{(n)}+
\frac{\xi}{2}\sum_{n\geq
1}\frac{n^2}{R^2}\bar{\chi}_a^{(n)}\frac{1}{\Box}\chi_a^{(n)}
+\frac{1}{2}\sum_{n\geq 1}\bar{\chi}_a^{(n)}\chi_a^{(n)} \right\}
\end{eqnarray*}

\begin{eqnarray*}
S_{matter} &=& \int\mbox{d}^8z \left\{\bar{\Phi}^{(0)}\Phi^{(0)} + \sum_{n\geq 1}(\bar{\Phi}^{(n)}\Phi^{(n)} + \Phi^{c(n)}\bar{\Phi}^{c(n)}) - \sum_{n\geq 1}\frac{n}{R}(\Phi^{c(n)}\Phi^{(n)}\delta(\bar{\theta}) + \bar{\Phi}^{(n)}\bar{\Phi}^{c(n)}\delta(\theta)) \right. \\
& & + g\left[ 2\bar{\Phi}^{(0)}V^{(0)}\Phi^{(0)} + 2\sum_{n\geq 1}(\bar{\Phi}^{(0)}V^{(n)}\Phi^{(n)} + \bar{\Phi}^{(n)}V^{(0)}\Phi^{(n)} + \bar{\Phi}^{(n)}V^{(n)}\Phi^{(0)}) \right. \\
& & \left. - 2\sum_{n\geq 1}\Phi^{c(n)}V^{(0)}\bar{\Phi}^{c(n)} + \sum_{n\geq 1}(\Phi^{c(n)}\chi^{(n)}\Phi^{(0)}\delta(\bar{\theta}) + \bar{\Phi}^{(0)}\bar{\chi}^{(n)}\bar{\Phi}^{c(n)}\delta(\theta))\right] \\
& & +g\left[\sqrt{2}\sum_{m,n\geq 1}\bar{\Phi}^{(m)}V^{(n)}(\Phi^{(m+n)}+\Phi^{(|m-n|)}) + \sqrt{2}\sum_{m,n\geq 1}\Phi^{c(m)}(V^{(m+n)}-V^{(|m-n|)})\bar{\Phi}^{c(n)} \right. \\
& & \left.\left. - \frac{1}{\sqrt{2}}\sum_{m,n\geq 1}(\Phi^{c(m)}\chi^{(n)}(\Phi^{(m+n)}-\Phi^{(|m-n|)})\delta(\bar{\theta}) + \bar{\Phi}^{(m)}\bar{\chi}^{(n)}(\bar{\Phi}^{c(m+n)}-\bar{\Phi}^{c(|m-n|)})\delta(\theta))\right]\right\}
\end{eqnarray*}

\begin{eqnarray*}
S_{brane} &=& \int\mbox{d}^8z \left\{\frac{\lambda_{ijk}}{6}\left[\Phi_i^{(0)}\Phi_j^{(0)}\Phi_k^{(0)} + 3\sqrt{2}\sum_{n\geq 1}\Phi_i^{(n)}\Phi_j^{(0)}\Phi_k^{(0)} + 6\sum_{m,n\geq 1}\Phi_i^{(0)}\Phi_j^{(m)}\Phi_k^{(n)} \right.\right. \\
& & \left. +2\sqrt{2}\sum_{m,n,p\geq 1}\Phi_i^{(m)}\Phi_j^{(n)}\Phi_k^{(p)}\right]\delta(\bar{\theta}) + \mbox{h.c.} \\
& & + \frac{\kappa_{ij}}{4}\left[L_i^{(0)}H_u^{(0)}L_j^{(0)}H_u^{(0)} + 2\sqrt{2}\sum_{n\geq 1}(L_i^{(n)}H_u^{(0)}L_j^{(0)}H_u^{(0)} + L_i^{(0)}H_u^{(n)}L_j^{(0)}H_u^{(0)}) \right. \\
& & + 4\sum_{m,n\geq 1}(L_i^{(m)}H_u^{(n)}L_j^{(0)}H_u^{(0)} + L_i^{(m)}H_u^{(0)}L_j^{(0)}H_u^{(n)}) \\
& & + 4\sqrt{2}\sum_{m,n,p\geq 1}(L_i^{(0)}H_u^{(m)}L_j^{(n)}H_u^{(p)} + L_i^{(m)}H_u^{(0)}L_j^{(n)}H_u^{(p)}) \\
& & \left. + 4\sum_{m,n,p,q\geq
1}L_i^{(m)}H_u^{(n)}L_j^{(p)}H_u^{(q)}\right]\delta(\bar{\theta}) +\mbox{h.c.}\left.\right\}~. 
\end{eqnarray*}
We kept the interaction terms involving only excited KK modes although
they are not relevant for the one-loop beta functions, since they can
be relevant to the renormalisation of the Kaluza-Klein masses and to
the mixing of the excited states.  We defined the 4D effective
couplings: $\displaystyle g=\frac{\tilde{g}}{\sqrt{\pi R}}$,
$\displaystyle \lambda=\frac{\tilde{\lambda}}{(\sqrt{\pi R})^3}$,
$\displaystyle \kappa=\frac{\tilde{\kappa}}{(\pi R)^2}$ 
.\\
\vspace{4mm}

The propagators are read off from the action (we choose the gauge $\xi=-1$)~:
\epsfig{file=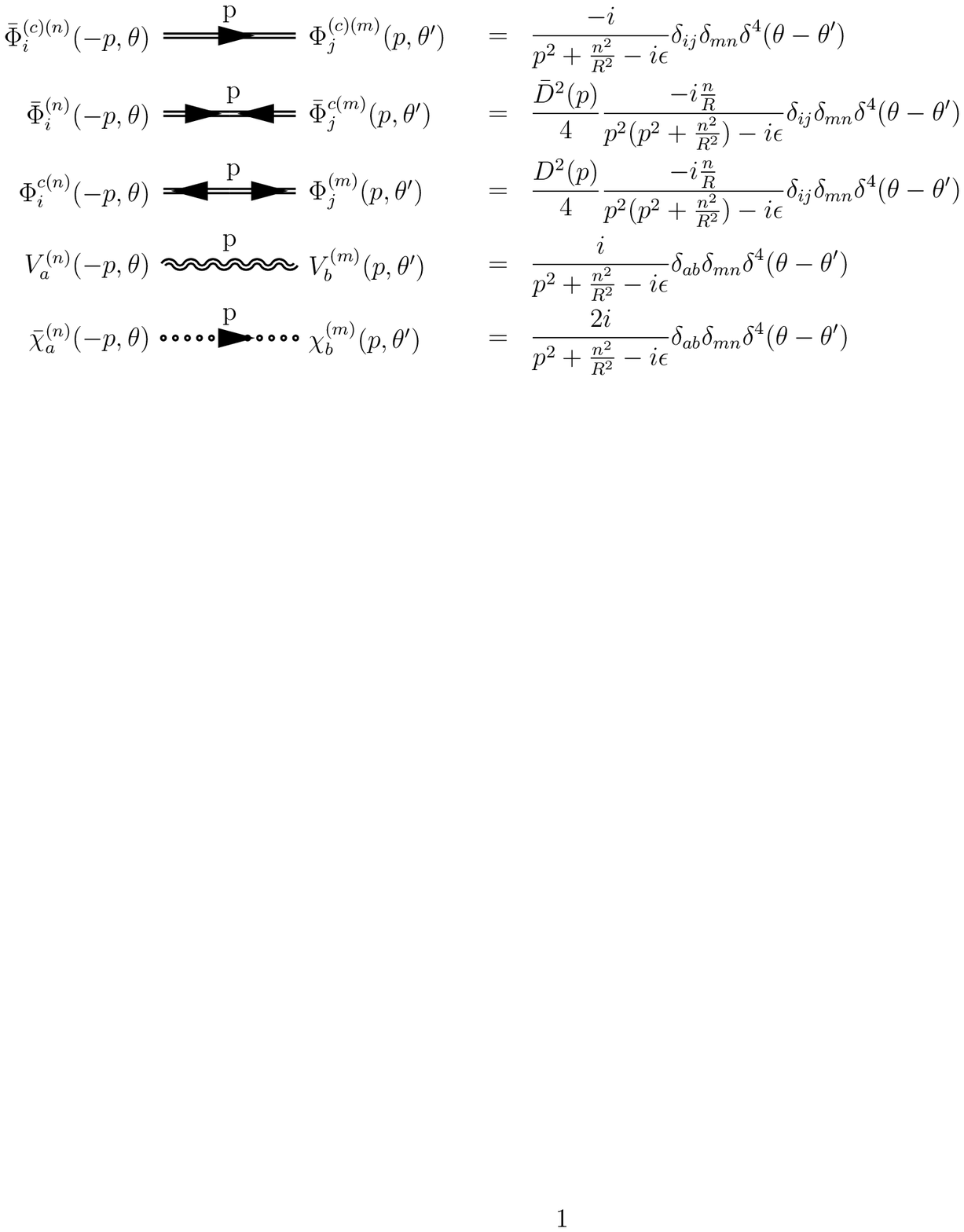}
and we derive the vertices that are relevant for our one loop calculation:

\hspace{-0.8cm}\epsfig{file=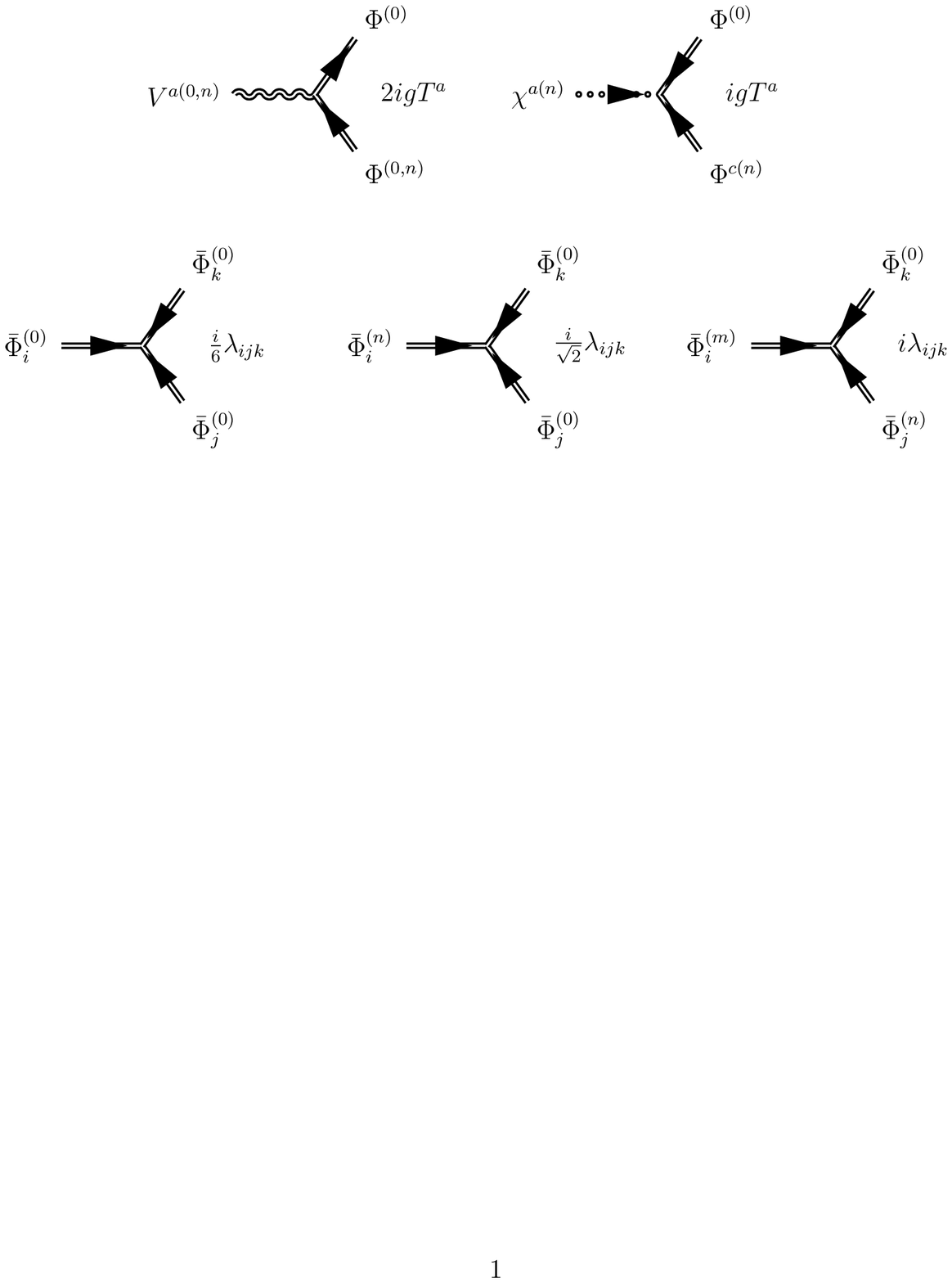}
\vspace{1cm}

We can do the same for the case where all superfields containing SM
fermions are restricted to the brane. The part of the action involving
only gauge and Higgs fields is not modified,
whereas the action for the superfields containing
the SM fermions splits into:
\begin{eqnarray*}
S_{brane} &=& \int\mbox{d}^8z\mbox{d}y\delta(y)\left\{\bar{\Phi}_i\Phi_i + 2\tilde{g}\bar{\Phi}_iV\Phi_i\right\} \\
&=& \int\mbox{d}^8z\left\{\bar{\Phi}_i\Phi_i + 2g\bar{\Phi}_iV^{(0)}\Phi_i + 2\sqrt{2}g\sum_{n\geq 1}\bar{\Phi}_iV^{(n)}\Phi_i\right\} \\
\end{eqnarray*}
\begin{eqnarray*}
S_{Yukawa} &=& \int\mbox{d}^8z\mbox{d}y\delta(y)\left\{\tilde{Y}_eE^cLH_d + \tilde{Y}_dD^cQH_d + \tilde{Y}_uU^cQH_u + \frac{1}{4}\tilde{\kappa}LH_uLH_u + \mbox{h.c.} \right\} \\
&=& \int\mbox{d}^8z\left\{ Y_eE^cLH_d^{(0)} + Y_dD^cQH_d^{(0)} + Y_uU^cQH_u^{(0)} + \frac{1}{4}\kappa LH_u^{(0)}LH_u^{(0)} \right.\\
&& + \sum_{n\geq 1}\sqrt{2}\left( Y_eE^cLH_d^{(n)} +Y_dD^cQH_d^{(n)} +Y_uU^cQH_u^{(n)} + \frac{1}{2}\kappa LH_u^{(n)}LH_u^{(n)}\right) \\
&& \left. + \sum_{m,n\geq 1}\frac{1}{2}\kappa LH_u^{(m)}LH_u^{(n)} + \mbox{h.c.}\right\}
\end{eqnarray*}
where we have written: $\displaystyle{Y_i=\frac{\tilde{Y_i}}{\sqrt{\pi
      R}}}$ et
$\displaystyle{\kappa=\frac{\tilde{\kappa}}{\pi R}}$.\\
The propagators and Feynman rules can be derived in the same way as
before.

\section{Kaluza-Klein integrals}
\label{integrals}
We will give here the major steps to the derivation of the relevant 
Kaluza-Klein
integrals and the computation of divergences following~\cite{DDG}. We will
not show finite terms, i.e. terms which vanish in the limit 
$\Lambda \to \infty$. For the calculation of the one-loop contributions
to the wave function renormalisation constants we need four types of two-point 
functions which we will discuss in turn. 

\subsection{Two excited KK modes with the same KK number}
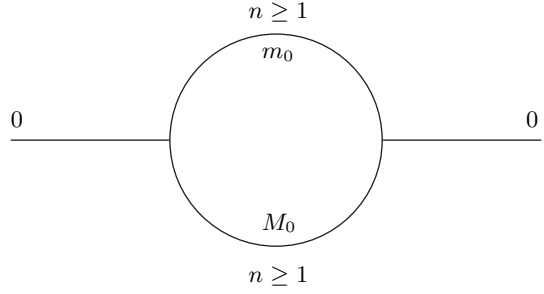
\begin{figure}
 \begin{center} 
 \begin{picture}(220,90)(0,-40)
\Line(0,0)(60,0)
\Line(140,0)(200,0)
\CArc(100,0)(40,0,360)
\Text(0,5)[lb]{$0$}
\Text(195,5)[lb]{$0$}
\Text(90,45)[lb]{$n\geq1$}
\Text(95,30)[lb]{$m_0$}
\Text(90,-55)[lb]{$n\geq1$}
\Text(95,-35)[lb]{$M_0$}
 \end{picture}
 \end{center}
\caption{One-loop diagram with two Kaluza-Klein states running in the loop 
restricted to have the same KK number}
 \label{bulkcontrib}
\end{figure}

The first case contains two excited KK modes, but their KK number is 
restricted to be the same. It is illustrated in Fig.~\ref{bulkcontrib} and 
arises typically from bulk interactions. For example, it enters the 
contribution from the one-loop correction containing one vector superfield
and one chiral superfield (which can be only Higgs for the model discussed in 
Sect.~\ref{secbrane} and Higgs or matter superfield for the model in 
Sec.~\ref{secbulk}). 

\noindent Let us start from the following general expression
\beq
\mathcal{K}=\sum_{n\geq 1}\int\frac{d^4k}{(2\pi)^4}\frac{1}{k^2-m_0^2-n^2/R^2}\frac{1}{k^2-M_0^2-n^2/R^2}~.
\eeq
After introducing 
Feynman parameters and performing a Wick rotation in the standard way, 
the resulting momentum integration can be re-expressed using a proper-time 
regularised form. This allows to introduce two cutoff parameters, 
$t_\mathrm{IR}$ and $t_\mathrm{UV}$ to treat infrared
and ultraviolet divergences, respectively. Assuming $M_0\sim m_0\ll R^{-1}$ 
the sum over KK states can be evaluated. We obtain
\beq
(16\pi^2)\mathcal{K}=\int^1_0dx\int^{t_{IR}}_{t_{UV}}\frac{dt}{t}\left[\frac{1}{2}\theta_3\left(\frac{it}{\pi
R^2}\right)-\frac{1}{2}\right]
\eeq
The $\theta_3$ function arising from the sum over KK states is defined as:
\beq 
\theta_3(x)=\sum_{n=-\infty}^{n=+\infty}e^{i\pi n^2x}~.
\eeq
As discussed in the appendix of Ref.~\cite{DDG} we can use the approximate form
of the $\theta_3$ function,
\beq
\theta_3\left(\frac{it}{\pi R^2}\right) \approx R \sqrt{\frac{\pi}{t}} 
\eeq
to evaluate the integral. This form of the $\theta_3$ function can be applied in 
general if $t_{\mathrm{IR}},\,t_{\mathrm{UV}}\ll R^{2}$,  but it gives also a 
very good approximation to the integral if $t_\mathrm{IR} \sim R^2$. We then 
obtain :  
\beqn
(16\pi^2)\mathcal{K}&\simeq& \int^1_0dx\int^{t_{\mathrm{IR}}}_{t_{\mathrm{UV}}}\frac{dt}{t}
\left(\frac{1}{2}\sqrt{\frac{\pi R^2}{t}}-\frac{1}{2}\right) \\
&=&\int^{t_{\mathrm{IR}}}_{t_{\mathrm{UV}}}dt\left(\frac{\sqrt{\pi}R}{2 t^{3/2}}-\frac{1}{2t}\right)
\eeqn
With the redefinitions $t_{\mathrm{UV}}=r\Lambda^{-2}$, $t_{\mathrm{IR}}=rR^{2}$ 
and $r = \pi/4$ the final result reads:
\beqn
(16\pi^2)\mathcal{K}&=&-2+2\Lambda R-\frac{1}{2}\log\Lambda^2R^2 \\
&\simeq&2\Lambda R-\frac{1}{2}\log\Lambda^2R^2~.\label{finalresult}
\eeqn
In the last line we have supposed that $\Lambda R \gg 1$. 

In Ref~\cite{Varin:2006wa}, the authors discussed another coherent cut-off 
regularisation scheme which allows for obtaining the preceeding result 
(\ref{finalresult}) naturally
without any rescaling of the cut-off when the KK tower is truncated at $\Lambda
R$.

\subsection{Two KK excitations with different KK numbers}

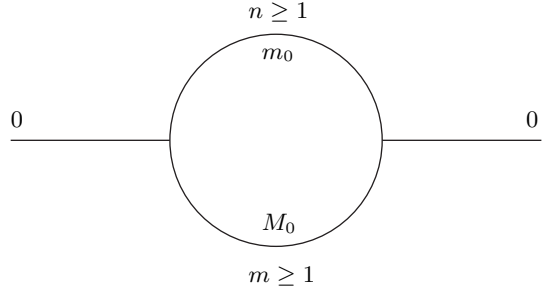
\begin{figure}[htbp!] 
 \begin{center} 
 \begin{picture}(220,80)(0,-40)
\Line(0,0)(60,0)
\Line(140,0)(200,0)
\CArc(100,0)(40,0,360)
\Text(0,5)[lb]{$0$}
\Text(195,5)[lb]{$0$}
\Text(90,45)[lb]{$n\geq1$}
\Text(95,30)[lb]{$m_0$}
\Text(90,-55)[lb]{$m\geq1$}
\Text(95,-35)[lb]{$M_0$}
 \end{picture}
 \end{center}
 \caption{One-loop diagram where two Kaluza-Klein states are
not constrained to be equal}
 \label{DoubleSumfig}
\end{figure}

We now perform the integral where two Kaluza-Klein states run in the loop and 
are not constrained to have the same KK number,
\beq
\mathcal{G}=\sum_{n,m\geq 1}\int\frac{d^4k}{(2\pi)^4}\frac{1}{k^2-m_0^2-n^2/R^2}\frac{1}{k^2-M_0^2-m^2/R^2}~.
\eeq
This type of integral arises only in the model discussed in 
Sec.~\ref{secbrane}, where all matter superfields propagate in the bulk. 
It appears in
connection with the Yukawa interactions restricted to the brane. 
This type of function is illustrated in 
Fig.~\ref{DoubleSumfig}.

\vspace{7mm}

Following the same steps as in the first case, we have:
\beqn
(16\pi^2)\mathcal{G}&=&\int^1_0dx\sum_{n,m \geq 1}\int\frac{dt}{t}e^{-\Delta_{nm} t} \nonumber\\
&=&\int^1_0dx\int^{t_{\mathrm{IR}}}_{t_{\mathrm{UV}}}\frac{dt}{t}\left[\frac{1}{2}\theta_3\left(\frac{itx}{\pi
R^2}\right)-\frac{1}{2}\right]\left[\frac{1}{2}\theta_3\left(\frac{it(1-x)}{\pi
R^2}\right)-\frac{1}{2}\right]~,
\eeqn
where
\beq
\Delta_{nm}=(m_0^2+n^2/R^2)x+(M_0^2+m^2/R^2)(1-x)\simeq n^2x/R^2 + m^2(1-x)/R^2~.
\eeq
In exactly the same way as in the previous section we obtain in this case
\beqn
(16\pi^2)\mathcal{G}&=&\pi\Lambda^2R^2-\pi+4-4\Lambda R+\frac{1}{4}\log\Lambda^2R^2 \\
&\simeq&\pi\Lambda^2R^2-4\Lambda R+\frac{1}{4}\log\Lambda^2R^2
\eeqn

\subsection{One KK mode running in the loop}
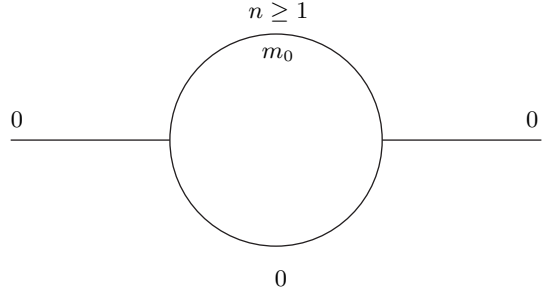
\begin{figure}[htbp!] 
 \begin{center} 
 \begin{picture}(220,80)(0,-40)
\Line(0,0)(60,0)
\Line(140,0)(200,0)
\CArc(100,0)(40,0,360)
\Text(0,5)[lb]{$0$}
\Text(195,5)[lb]{$0$}
\Text(90,45)[lb]{$n\geq1$}
\Text(95,30)[lb]{$m_0$}
\Text(100,-55)[lb]{$0$}
 \end{picture}
 \end{center}
 \caption{One-loop diagram with only one Kaluza-Klein state in the loop.}
 \label{SimpleSumfig}
\end{figure}

The third type of integral contains one zero mode and one excited KK mode:
\beq
\mathcal{H}=\sum_{n\geq 1}\int\frac{d^4k}{(2\pi)^4}\frac{1}{k^2-m_0^2-n^2/R^2}\frac{1}{k^2-M_0^2}~.
\eeq
It is illustrated in Fig.~\ref{SimpleSumfig}.
The evaluation proceeds in exactly the same way as before. We obtain
\beqn
(16\pi^2)\mathcal{H}&=&\int^1_0dx\sum_{n\geq 1}\int\frac{dt}{t}e^{-n^2x/R^2 t} \nonumber\\
&=&\int^1_0dx\int^{t_{\mathrm{IR}}}_{t_{\mathrm{UV}}}\frac{dt}{t}\left[\frac{1}{2}\theta_3\left(\frac{itx}{\pi
R^2}\right)-\frac{1}{2}\right]
\eeqn
which gives upon performing the same approximations as before
\beqn
(16\pi^2)\mathcal{H}&=&4\Lambda R-4-\frac{1}{2}\log\Lambda^2 R^2 \\
&\simeq&4\Lambda R-\frac{1}{2}\log\Lambda^2 R^2.
\eeqn
\subsection{Only zero modes running in the loop}
We recall the result of a loop with only zero modes,
\beq
\mathcal{I}=\int\frac{d^4k}{(2\pi)^4}\frac{1}{k^2-m_0^2}\frac{1}{k^2-M_0^2}~,
\eeq
which reads
\beq
(16\pi^2)\mathcal{I}=\int^1_0dx\int\frac{dt}{t}e^{-t(m_0^2x+M_0^2(1-x))}\simeq \log\Lambda^2 R^2.
\eeq 
\section{Renormalisation constants}
Within this section we summarise the explicit expressions for all wave 
function renormalisation constants needed to compute the beta functions.
In this section we will display the diagrams entering in the one-loop 
renormalisation of the Yukawa couplings and their values, in order to 
deduce the wave functions renormalisation and the beta functions. 
Extensive use is made of the integrals calculated in the last section.

\subsection{Matter fields propagating in the bulk}
\label{zbulk}
We have 5 types of diagrams:

\hspace{-1.1cm}\epsfig{file=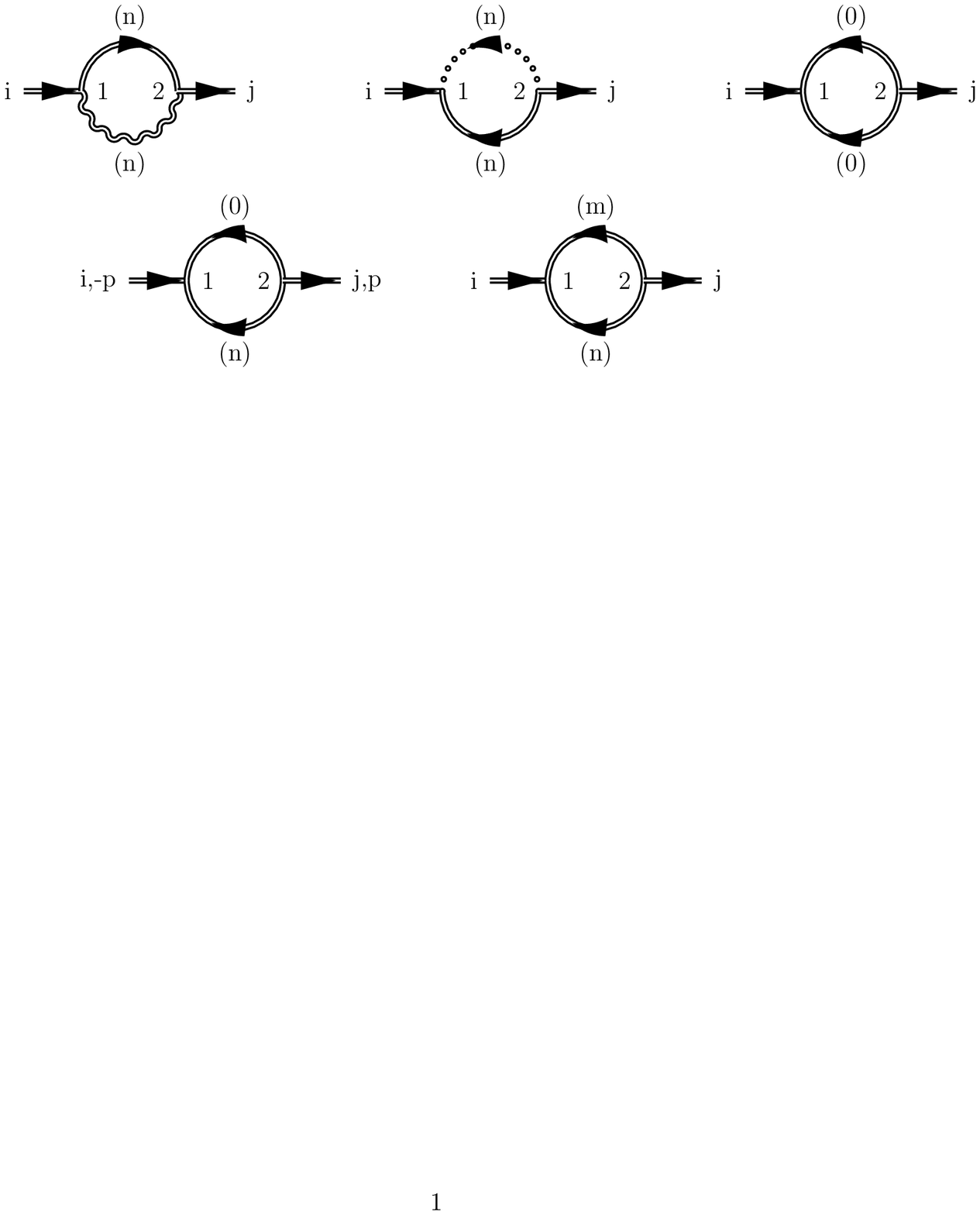}
\vspace{5mm}

We provide some steps of the calculation for the first diagram and give the result for the four others~:

\begin{eqnarray*}
\delta Z_{ij}^{(1)} &=& -4g^2(T^aT^a)_{rs}\delta_{ij}\sum_{n\geq 0}\int\frac{\mbox{d}^4k}{(2\pi)^4}\mbox{d}^4\theta_1\mbox{d}^4\theta_2\frac{i\delta^4_{12}}{2(k^2+\frac{n^2}{R^2})} \\
&& \qquad\qquad\qquad\qquad\qquad \times\frac{1}{16}\bar{D}_1^2 D_2^2\frac{-i\delta^4_{12}}{(k+p)^2+\frac{n^2}{R^2}}\phi_i^{r(0)}(-p,\theta_1)\bar{\phi}_j^{s(0)}(p,\theta_2) \\
&=& -i2g^2C_2(R)\delta_{ij}^{rs}\left(\mathcal{K}+\int\frac{\mbox{d}^4k}{(2\pi)^4}\frac{1}{k^2(p+k)^2}\right)\int\mbox{d}^4\theta_1\phi_i^{r(0)}(-p,\theta_1)\bar{\phi}_j^{s(0)}(p,\theta_1) \\
&=& i\frac{-2g^2C_2(R)\delta_{ij}^{rs}}{16\pi^2}\left(2\Lambda R+\log(\Lambda R)\right)\int\mbox{d}^4\theta_1\phi_i^{r(0)}(-p,\theta_1)\bar{\phi}_j^{s(0)}(p,\theta_1)
\end{eqnarray*}

\begin{eqnarray*}
\delta Z_{ij}^{(2)} &=& i\frac{-2g^2C_2(R)\delta^{rs}_{ij}}{16\pi^2}\left(2\Lambda R-\log(\Lambda R)\right)
\qquad
\delta Z_{ij}^{(3)} = i\frac{\lambda_{ikl}\lambda^*_{jkl}\delta_{rs}}{16\pi^2}\log(\Lambda R) \\
\delta Z_{ij}^{(4)} &=& i\frac{2\lambda_{ikl}\lambda^*_{jkl}\delta_{rs}}{16\pi^2}\left(4\Lambda R - \log(\Lambda R)\right)
\qquad
\delta Z_{ij}^{(5)} = i\frac{\lambda_{ikl}\lambda^*_{jkl}\delta_{rs}}{16\pi^2}\left(2\pi(\Lambda R)^2 - 8\Lambda R + \log(\Lambda R) \right)
\end{eqnarray*}
We only displayed the integral over the $\theta$ coordinate for the first 
contribution, omitting it in the others.\\

Summing the different contributions, every subdominant divergence disappears for both the gauge and the Yukawa contribution and we obtain:
\beq
-(16\pi^2)\,\delta Z^{5D}_{\Phi}=
\left(-8\sum_{n=1}^{N_{g}}g_n^2C_2(R_n^{(i)})\delta_{ij}\right)\Lambda R
+\left(2\pi\sum_{k,l=1}^{N_{\Phi}}\lambda^*_{ikl}\lambda_{jkl}\right)\Lambda^2R^2~.
\eeq 

Applying it to the matter and Higgs superfields :
\beqn
-(16\pi^2)\,\delta Z_{H_u}&=&12\pi \mathrm{Tr}(Y_u^{\dag}Y_u)\Lambda^2R^2-(\frac{6}{5}g_1^2+6g_2^2)\Lambda R \\
-(16\pi^2)\,\delta Z_{H_d}&=&4\pi[3 \mathrm{Tr}(Y_d^{\dag}Y_d)+\mathrm{Tr}(Y_e^{\dag}Y_e)]\Lambda^2R^2-(\frac{6}{5}g_1^2+6g_2^2)\Lambda R \\
-(16\pi^2)\,\delta Z_{L}&=&4\pi(Y_e^{\dag}Y_e)\Lambda^2R^2-(\frac{6}{5}g_1^2+6g_2^2)\Lambda R \\
-(16\pi^2)\,\delta Z_{E^C}&=&8\pi(Y_e^*Y_e^T)\Lambda^2R^2-(\frac{24}{5}g_1^2)\Lambda R\\
-(16\pi^2)\,\delta Z_{D^c}&=&8\pi(Y_d^*Y_d^T)\Lambda^2R^2-(\frac{8}{15}g_1^2+\frac{32}{3}g_3^2)\Lambda R \\
-(16\pi^2)\,\delta
Z_{Q}&=&4\pi(Y_u^{\dag}Y_u+Y_d^{\dag}Y_d)\Lambda^2R^2-(\frac{2}{15}g_1^2+6g_2^2+\frac{32}{3}g_3^2)\Lambda R \\
-(16\pi^2)\,\delta Z_{U^c}&=&8\pi(Y_u^*Y_u^T)\Lambda^2R^2-(\frac{32}{15}g_1^2+\frac{32}{3}g_3^2)\Lambda R
\eeqn
from which we deduce the Yukawa beta functions :
\beqn
\beta_{Y_d}&=&-\frac{1}{2}\Lambda\frac{\partial}{\partial\Lambda}(\delta Z_{D^c}^{T}Y_d+Y_d\delta
Z_{Q}+Y_d\delta Z_{H_d}) \nonumber\\
\beta_{Y_u}&=&-\frac{1}{2}\Lambda\frac{\partial}{\partial\Lambda}(\delta Z_{U^c}^{T}Y_u+Y_u\delta
Z_{Q}+Y_u\delta Z_{H_u}) \nonumber\\
\beta_{Y_d}&=&-\frac{1}{2}\Lambda\frac{\partial}{\partial\Lambda}(\delta Z_{E^c}^{T}Y_e+Y_e\delta
Z_{L}+Y_e\delta Z_{H_d}) \nonumber
\eeqn

\subsection{Matter fields restricted to the brane}
\label{zbrane}
There again we show all 4 diagrams contributing for the 3 generations of flavour:

\epsfig{file=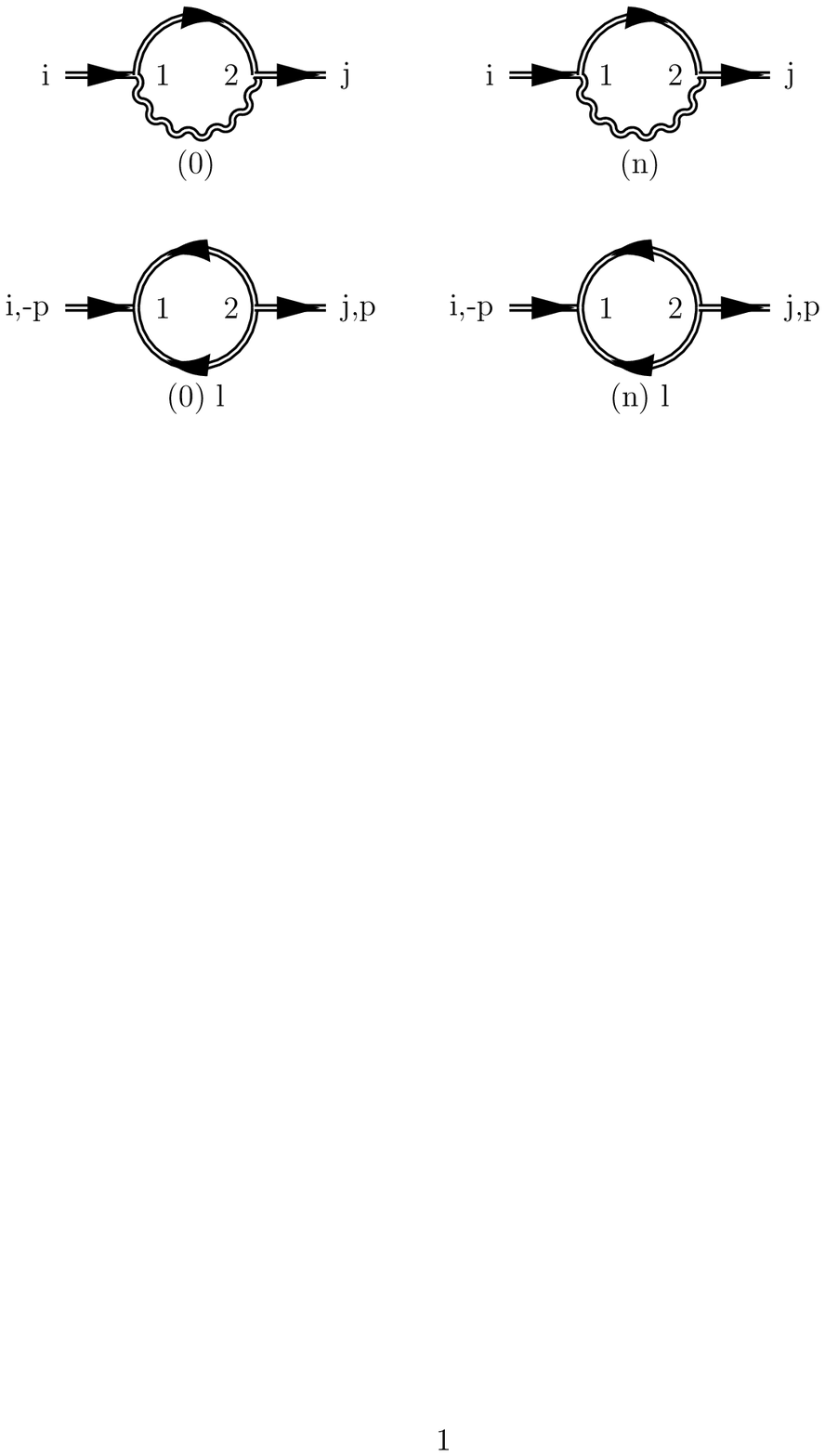}

\begin{eqnarray*}
\delta Z_{ij}^{(1)} &=& i\frac{-4g^2C_2(R)\delta_{ij}^{rs}}{16\pi^2}\log(\Lambda R)
\qquad
\delta Z_{ij}^{(2)} = i\frac{-8g^2C_2(R)\delta_{ij}^{rs}}{16\pi^2}\left( 4\Lambda R - \log(\Lambda R)\right) \\
\delta Z_{ij}^{(3)} &=& i\frac{\lambda_{ikl}\lambda^*_{jkl}}{16\pi^2}\log(\Lambda R)
\qquad\qquad\quad
\delta Z_{ij}^{(4)} = i\frac{\lambda_{ikl}\lambda^*_{jkl}}{16\pi^2}\left(4\Lambda R - \log(\Lambda R)\right)
\end{eqnarray*}
The sum gives~:
\beq
\delta Z^L_{ij}=-\frac{1}{16\pi^2}\left[ -16\Lambda Rg^2C_2(R)\delta_{ij} + 4\Lambda R\lambda_{ikl}\lambda^*_{jkl} \right]
\eeq
As for the Higgses, the gauge diagrams are those the previous case and for $\delta Z$ we collect~:
\beq
\delta Z^H = -\frac{1}{16\pi^2}\left[ -8\Lambda Rg^2C_2(R)+ 2\log(\Lambda R)Tr(Y_iY_i^\dagger) \right]
\eeq
It is straightforward to deduce the following renormalisation constants for the matter and Higgs superfields :
\beqn
-(16\pi^2)\,\delta Z_{H_u}&=&6\mathrm{Tr}(Y_u^{\dag}Y_u)\log\Lambda R-(\frac{6}{5}g_1^2+6g_2^2)\Lambda R\\
-(16\pi^2)\,\delta
Z_{H_d}&=&[6\mathrm{Tr}(Y_d^{\dag}Y_d)+2\mathrm{Tr}(Y_e^{\dag}Y_e)]\log\Lambda R-(\frac{6}{5}g_1^2+6g_2^2)\Lambda R\\
-(16\pi^2)\,\delta Z_{L}&=&[8 (Y_e^{\dag}Y_e)-\frac{12}{5}g_1^2-12g_2^2]\Lambda R \\
-(16\pi^2)\,\delta Z_{E^C}&=&[16 (Y_e^*Y_e^T)-\frac{48}{5}g_1^2]\Lambda R\\
-(16\pi^2)\,\delta Z_{D^c}&=&[16 (Y_d^*Y_d^T)-\frac{16}{15}g_1^2-\frac{64}{3}g_3^2]\Lambda R \\
-(16\pi^2)\,\delta
Z_{Q}&=&[8 (Y_u^{\dag}Y_u+Y_d^{\dag}Y_d)-\frac{4}{15}g_1^2-12g_2^2-\frac{64}{3}g_3^2]\Lambda R \\
-(16\pi^2)\,\delta Z_{U^c}&=&[16(Y_u^*Y_u^T)-\frac{64}{15}g_1^2-\frac{64}{3}g_3^2]\Lambda R
\eeqn

\end{document}